\documentclass[1p]{elsarticle}
\date{March 26, 2025}

\usepackage{amssymb}
\usepackage{amsmath}
\usepackage{multirow}
\usepackage{tikz}
\usepackage[numbers]{natbib}

\newcommand{\ovalbox}[1]{\tikz[baseline=(char.base)]{
            \node[shape=rectangle,draw,inner sep=2pt] (char) {#1};}}

\newcommand{\odashbox}[1]{\tikz[baseline=(char.base)]{
            \node[shape=rectangle,draw,dotted,line width=0.75pt,inner sep=2pt] (char) {#1};}}

\usepackage{hyperref}
\hypersetup{
    colorlinks,
    citecolor=black,
    filecolor=black,
    linkcolor=black,
    urlcolor=black
}
\usepackage{booktabs}

\usepackage{siunitx}
\usepackage{longtable}
\usepackage{lscape}
\usepackage{subcaption}
\usepackage{placeins}
\usepackage[acronym,nonumberlist]{glossaries}

\journal{Aerospace Science and Technology}

\newglossaryentry{atc}{
    name={ATC},
    description={Air traffic control}
}
\newglossaryentry{atco}{
    name={ATCO},
    description={Air traffic control officer}
}
\newglossaryentry{bada}{
    name={BADA},
    description={Base of Aircraft Data}
}
\newglossaryentry{bic}{
    name={BIC},
    description={Bayesian information criterion}
}
\newglossaryentry{bnn}{
    name={BNN},
    description={Bayesian neural network}
}
\newglossaryentry{cas}{
    name={CAS},
    description={Calibrated airspeed. Equivalent to true airspeed when flying at sea level under International Standard Atmosphere conditions}
}
\newglossaryentry{cdf}{
    name={CDF},
    description={Cumulative distribution function}
}
\newglossaryentry{cdo}{
    name={CDO},
    description={Continuous descent operations}
}
\newglossaryentry{esf}{
    name={ESF},
    description={Energy share factor}
}
\newglossaryentry{fl}{
    name={FL},
    description={Flight level}
}
\newglossaryentry{fir}{
    name={FIR},
    description={Flight information region}
}
\newglossaryentry{fpca}{
    name={fPCA},
    description={Functional principal component analysis}
}
\newglossaryentry{gcs}{
    name={GCS},
    description={Geographical coordinate space}
}
\newglossaryentry{ga}{
    name={GA},
    description={General aviation}
}
\newglossaryentry{gmm}{
    name={GMM},
    description={Gaussian mixture model}
}
\newglossaryentry{ias}{
    name={IAS},
    description={Indicated airspeed. The \gls{cas} measured by an aircraft's sensors, subject to measurement and position errors}
}
\newglossaryentry{icao}{
    name={ICAO},
    description={International Civil Aviation Organization}
}
\newglossaryentry{isa}{
    name={ISA},
    description={International standard atmosphere}
}
\newglossaryentry{ks}{
    name={KS},
    description={Kolmogorov-Smirnov (distance)}
}
\newglossaryentry{kl}{
    name={KL},
    description={Kullback-Leibler (divergence)}
}
\newglossaryentry{lstm}{
    name={LSTM},
    description={Long short-term memory}
}
\newglossaryentry{mad}{
    name={MAD},
    description={Mean absolute deviation}
}
\newglossaryentry{mae}{
    name={MAE},
    description={Mean absolute error}
}
\newglossaryentry{map}{
    name={MAP},
    description={Maximum a posteriori}
}
\newglossaryentry{mse}{
    name={MSE},
    description={Mean squared error}
}
\newglossaryentry{ml}{
    name={ML},
    description={Machine learning}
}
\newglossaryentry{nf}{
    name={NF},
    description={Normalizing flow}
}
\newglossaryentry{rocd}{
    name={ROCD},
    description={Rate of climb/descent}
}
\newglossaryentry{tas}{
    name={TAS},
    description={True airspeed. Aircraft speed relative to the airmass through which it is flying}
}
\newglossaryentry{pca}{
    name={PCA},
    description={Principal component analysis}
}
\newglossaryentry{pdf}{
    name={PDF},
    description={Probability density function}
}
\newglossaryentry{piml}{
    name={PIML},
    description={Physics-informed machine learning}
}
\newglossaryentry{pp}{
    name={PP},
    description={Power predictive}
}
\newglossaryentry{TP}{
    name={TP},
    description={Trajectory Prediction}
}

\makeglossaries

\begin{document}

\begin{frontmatter}

\title{Probabilistic Simulation of Aircraft Descent via a Physics-Informed Machine Learning Approach}

\author[1]{Hodgkin, A.}
\author[1]{Pepper, N.}
\author[2,3,4]{Thomas, M.}

\affiliation[1]{organization={The Alan Turing Institute},
            addressline={British Library},
            city={London},
            postcode={NW1 2DB}, 
            country={UK}}

\affiliation[2]{organization={NATS},
            addressline={Whitely}, 
            city={Fareham},
            postcode={PO15 7FL}, 
            country={UK}}
\affiliation[3]{organization={Queen Mary University},
            addressline={Bancroft Rd},
            city={London},
            postcode={E1 4DG}, 
            country={UK}}
\affiliation[4]{organization={University of Cambridge},
            city={Cambridge},
            postcode={CB2 1TN}, 
            country={UK}}

\begin{abstract}
This paper presents a method for generating probabilistic descent trajectories in simulations of real-world airspace. A dataset of {116,066} trajectories harvested from Mode S radar returns in UK airspace was used to train and test the model. Thirteen aircraft types with varying performance characteristics were investigated. It was found that the error in the mean prediction of time to reach the bottom of descent for the proposed method was less than that of the the Base of Aircraft Data (\gls{bada}) model by a factor of 10. Furthermore, the method was capable of generating a range of trajectories that were similar to the held out test dataset when analysed in distribution. The proposed method is hybrid, with aircraft drag and calibrated airspeed functions generated probabilistically to parameterise the \gls{bada} equations, ensuring the physical plausibility of generated trajectories. 
\end{abstract}

\begin{keyword}
Air Traffic Management \sep Trajectory Prediction \sep Probabilistic Machine Learning \sep Normalizing Flows
\end{keyword}

\end{frontmatter}

\printglossary

\section{Introduction}
Air Traffic Control (\gls{atc}) issues instructions to aircraft in order to prevent collisions by ensuring adequate separation between aircraft, as well as enabling the expeditious and orderly flow of air traffic \cite{Mats_part1}. Other than a recent dip due to Covid-19, the number of flights flown annually has increased steadily each year, a trend which is expected to continue \cite{easr_2022}. This has led to a requirement to increase capacity and reduce climate emissions, without compromising on safety, that has driven large modernisation initiatives within Air Traffic Management Systems (see, e.g. \cite{sesar}). 

\textcolor{black}{Meeting these requirements requires the development of tools capable of providing decision support, improving workload forecasting and planning, enabling more fuel efficient procedures and systems, optimising route networks and improving network traffic prediction \cite{Sesar_sol}. Trajectory prediction (\gls{TP}) underpins these tasks. However, \gls{TP} is complicated by the presence of significant epistemic uncertainties which can arise due to unknown aircraft mass, wind conditions, aircraft performance settings, and pilot behaviours \cite{Lymperopoulos2010Sequential}. There is therefore value in developing a \gls{TP} model that can be used to address the above challenges by accurately modelling the uncertainty inherent in the system. An ideal probabilistic \gls{TP} model would have three features:}

\begin{enumerate}
    \item The capability of generating probabilistic trajectories that follow a similar distribution to what is observed in real-world operations. \\
 \item A mean trajectory prediction that could be calibrated to observed trajectories within a specific airspace. \\
 \item A guarantee that generated trajectories are physically plausible. For instance, aircraft in descent must decrease altitude monotonically within sensible performance limits for that aircraft type.  
\end{enumerate}

The current state of the art \gls{TP} model used in \gls{atc} is the Base of Aircraft Data (\gls{bada}) model \cite{nuic2010bada}. It is a total-energy model that takes parameters including the aircraft mass, speed profile, wind and atmospheric conditions, mode of flight, and aircraft reference data. It can be applied to airspaces worldwide. However BADA is a simplified model, which assumes generic operational procedures which can vary between operators, routes and airspaces. This can lead to significant model mis-specification between \gls{bada} and observed trajectories within a specific airspace, particularly in descent. Therefore, while \gls{bada} ensures that trajectories are physically plausible, as a deterministic model that is globally calibrated, it does not satisfy the first two of the desired features. 

There has been a great deal of recent research that has sought to develop more accurate \gls{TP} models that are specific to a particular airspace, often using machine learning (\gls{ml}) techniques to learn from large datasets of operational data. Examples of such techniques include: Bayesian Neural Networks \cite{pang2021data, pepper_sector}, Long Short-Term Memory Networks \cite{shi20204, tran2022aircraft}, Convolutional Neural Networks \cite{ma2020hybrid, krauth2023deep}, Generative Adversarial Networks \cite{wu2022long}, and Gaussian mixture models \cite{barratt2018learning}. Physics-informed machine learning (\gls{piml}) offers a range of methodologies to combine physics with machine learning to improve model outputs \cite{willard2022integrating}. For instance, it is possible to bias the learning of such models towards physically plausible solutions \cite{bastas2020data,pang2020conditional}. However, inductive bias in training does not guarantee that generated trajectories will be plausible, especially if the \gls{ml} model begins extrapolating away from its training dataset. Probabilistic models may be formulated such that monotonicity constraints are enforced (see, e.g. \cite{pepper2023probabilistic}\cite{riihimaki2010gaussian}). However, such a formulation can limit the flexibility of the model to fit the data.



\begin{figure}[ht!]
    \centering
    \includegraphics[width=0.7\linewidth]{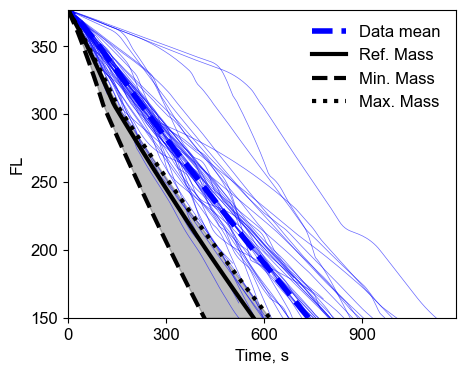}
    \caption{Effect of varying the mass in the \gls{bada} equation compared with true historical data for the B738. The data mean is computed from 41,011 trajectories and a randomly selected sample of 50 are shown here.}
    \label{fig:bada_mass}
\end{figure}

An alternative method for ensuring the physical plausibility of trajectories generated by an \gls{ml} method is to use data to update parameters in the \gls{bada} equations. Typically this is done through inferring the aircraft mass parameter in the \gls{bada} equations. This strategy can be very effective for improving trajectory predictions in climb, as demonstrated in Thipphavong et al. \cite{thipphavong2013adaptive} and Alligier et al. \cite{alligier2013learning}. However, in descent there may not be sufficient flexibility in the corrected model to fit real-world trajectory data using only the aircraft mass. Accurate \gls{TP} of aircraft descent is notoriously difficult, primarily because of the effect of airspace and airline procedures on the performance characteristics of aircraft in descent, making the descent phase less predictable than when in climb. This is demonstrated in Figure~\ref{fig:bada_mass}, where a selection of B738 descents are plotted in blue, with the dashed blue line indicating the mean of the data. The solid black line indicates the \gls{bada} trajectory using the reference mass for the B738 in the model. Dashed lines indicate the trajectories corresponding to the maximum and minimum mass in \gls{bada}, with light grey lines indicating trajectories corresponding to intermediate mass values. Figure~\ref{fig:bada_mass} shows that there is greater variation within a dataset of descending aircraft than can be accounted for purely by adjusting the aircraft mass parameter in \gls{bada}. 

Greater flexibility can be incorporated into the correction to \gls{bada} through the functional modelling of terms within the \gls{bada} equation, rather than through updating scalar parameters such as the aircraft mass. Functional Principal Component Analysis (\gls{fpca}) has been used by Nicol \cite{nicol2013functional} as a means of detecting outliers in a set of trajectories. Pepper and Thomas \cite{pepper2023learning} used \gls{fpca} to provide a convenient basis on which to model the thrust function in \gls{bada}. In so doing they were able to develop a probabilistic model that could capture the range of aircraft performance in a large dataset of climbing aircraft. In particular, this paper expands this methodology in a number of important respects, to account for the more complex task of simulating aircraft in descent: 

\begin{enumerate}
    \item \textbf{Combined {\gls{rocd}} and speed modelling}. The tradeoff between airspeed and rate of descent is an important feature of modelling aircraft descents. Pilots will aim to descend in idle thrust, without applying speed brakes. For environmental reasons, \gls{atc} providers increasingly aim to descend aircraft continuously, minimising time spent in holds and without level-offs at intermediate flight levels. This paradigm is referred to as Continuous Descent Operations (\gls{cdo}) with aircraft  following neutral descent profiles \cite{cdo}. Accurate modelling of a descent therefore requires modelling of speed as a well as altitude over time. In Pepper and Thomas \cite{pepper2023learning} the correction was applied only to the aircraft thrust, using the reference speeds from the \gls{bada} model. \\
    \item \textbf{Simulation across upper airspace}. A further limitation of Pepper and Thomas \cite{pepper2023learning} was that the thrust correction could only be applied for a range of altitudes that were spanned by every trajectory in the dataset. This paper introduces a ‘gappy’ algorithm for \gls{fpca} that accounts for trajectories in the training dataset that do not span the complete altitude range. This allows the algorithm presented here to be applied to a wider altitude range, from \gls{fl}s 150 to 398, compared to 150 to 325 in Pepper and Thomas \cite{pepper2023learning} and makes the algorithm more data efficient as fewer trajectories are discarded in data cleaning. \\
    \item \textbf{High fidelity probabilistic modelling}. The choice of probabilistic model in the latent space of \gls{fpca} coefficients is investigated more thoroughly in this paper. The effectiveness of Gaussian Mixture Models (\gls{gmm}s) and Normalising Flow (\gls{nf}) models \cite{papamakarios2021normalizing} for generating samples in this latent space are assessed. \\
    \item \textbf{Suggested metrics with relevance to {\gls{atc}}}. Assessing the capability of probabilistic trajectory models requires new metrics for identifying how closely the distribution of generated trajectories match a held out dataset. This paper suggests a set of metrics through which the quality of generated trajectories might be assessed. \\
\end{enumerate}

The remainder of the paper is structured as followed:  Section \ref{sec:method} details the method, including a short description of the \gls{bada} equations, \gls{fpca} and methods for learning the joint distribution of \gls{fpca} weights, Section \ref{sec:data} discusses the data preparation, and Section \ref{sec:results} proposes metrics for assessing the quality of the trajectory generation and evaluates trajectories sampled from the models. Conclusions are presented in Section \ref{concs}. 

\section{Probabilistic modelling of aircraft descents} \label{sec:method}
This section outlines the proposed probabilistic model for generating realistic descent trajectories. There are two components: a probabilistic model for aircraft drag and calibrated airspeed (\gls{cas}) and the \gls{bada} equations. When sampled, the probabilistic model produces drag and \gls{cas} as functions of geodetic altitude. These samples parameterise the \gls{bada} equations, ensuring that generated descent profiles will be physically plausible. This section briefly outlines the \gls{bada} model, before discussing the details of the probabilistic model for drag and \gls{cas} functions.

\subsection{The BADA model}
The \gls{bada} model is a set of physics-based equations and aircraft specific parameters for predicting trajectories. The total-energy equation, which relates the change in potential and kinetic energy to the work done on the aircraft, is defined in Nuic et al. \cite{nuic2010bada} as:
\begin{equation} 
    (T_{HR}-D)V_{TAS}=m\bigg(g_0\frac{dh}{dt} + V_{TAS} \frac{dV_{TAS}}{dt}\bigg), \label{eq:bada}
\end{equation}
where $T_{HR}$ is the aircraft thrust parallel to the velocity vector, $D$ is the drag, $V_{TAS}$ is the aircraft true airspeed (\gls{tas}), $m$ is the aircraft mass, $h$ is the aircraft geodetic altitude, and $g_0$ is the gravitational constant. \gls{bada} provides equations for $T_{HR}$ and $D$ as functions of altitude and flight condition. There is also a dependence on aircraft mass in $D$. 

\begin{figure}
    \centering
    \includegraphics[width=1\linewidth]{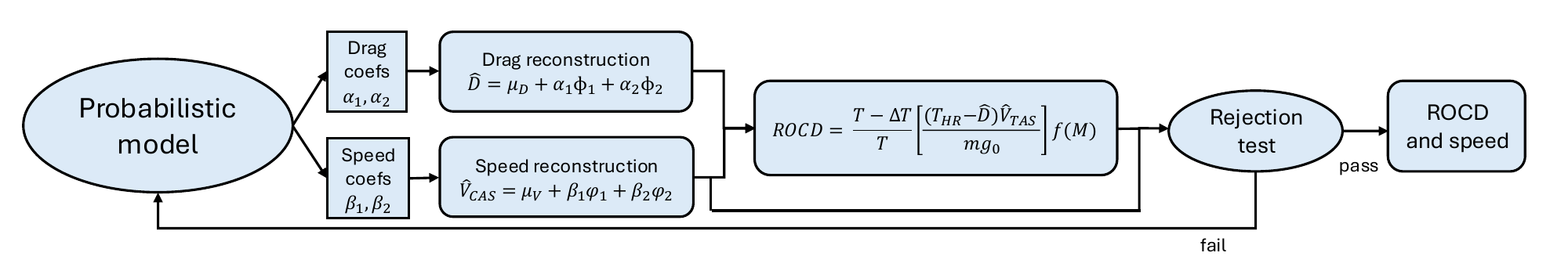}
    \caption{Schematic diagram showing the \gls{piml} model process. Oval blocks indicate operations, box blocks indicate latent space samples and curved blocks indicate functional samples.}
    \label{fig:modelschematic}
\end{figure}

\gls{bada} uses the International Standard Atmosphere (\gls{isa}) model to model variations of pressure and air density with $h$, allowing formulae for converting \gls{tas} to \gls{cas}, denoted $V_{CAS}$, to be defined. In order to solve \eqref{eq:bada} it is necessary to model the tradeoff between kinetic and potential energy as a function of altitude. This function is referred to as the energy share factor (\gls{esf}) and is multi-valued. The exact value used depends on whether the aircraft is above or below transition altitude and above or below the tropopause. The \gls{bada} reference manual details the steps by which the rate of climb/descent (\gls{rocd}) may be derived as a function of the \gls{esf} \cite{nuic2010user}:

\begin{equation}
    \frac{dh}{dt} = \frac{T-\Delta T}{T} \Big[ \frac{(T_{HR}-D)V_{TAS}}{mg_0} \Big] f(M), \label{eq:bada_rocd}
\end{equation}
where $\frac{dh}{dt}$ denotes the \gls{rocd} and $f(\cdot)$ the \gls{esf}, with the latter defined as a function of the Mach number, $M$. Additionally, $T$ represents air temperature in the \gls{isa} model and $\Delta T$ a temperature correction that may be applied to \gls{bada}, although in what follows this is set to 0. The model proposed in this paper generates functions for $D$ and $V_{CAS}$ as a function of $h$. The generated \gls{cas} is converted to \gls{tas} using the existing conversions in \gls{bada}. The algorithm to determine the \gls{esf} value is unchanged. Constants in the model, such as the aircraft mass, are set to their nominal \gls{bada} values. Formulae in the \gls{bada} model are unchanged, aside from those formula setting drag and \gls{cas} values. {Figure~\ref{fig:modelschematic}} is a schematic illustrating the methodology described here. 

\subsection{Probabilistic modelling of drag and CAS}\label{sec:drag_cas}
A probabilistic model generates drag and \gls{cas} profiles as functions of $h$. This model is developed from a dataset of drag and \gls{cas} data harvested from Mode S radar data (see, e.g. \cite{orlando1989mode}). Indicated airspeed (\gls{ias}) is returned directly from aircraft and can be used as measurements of $V_{CAS}$ subject to instrument noise. Aircraft drag is not measured by the aircraft but can be inferred using the \gls{bada} model and measurements of \gls{ias}, Mach number, and \gls{rocd} returned by the aircraft through a rearrangement of \eqref{eq:bada_rocd}: 
\begin{equation}
    \hat{D}=T_{HR} - \frac{T}{T-\Delta T} \bigg[ \frac{(dh/dt)  m g_0}{f(M) V_{TAS}} \bigg], \label{eq:infer}
\end{equation}
where $\hat{D}$ is used to denote that drag is a modelled, rather than measured, quantity. Mode S data and \eqref{eq:infer} can be used to generate a dataset of $V_{CAS}$ and $\hat{D}$ at geodetic altitude $h$ for every radar blip in a trajectory. Functional Principal Component Analysis (\gls{fpca}) provides a convenient basis with which to represent variations of aircraft drag and \gls{cas} with altitude and decompose trajectory data onto a lower dimensional latent space, with a representation of the form:

\begin{subequations} \label{eq:fpca}
\begin{equation}
    \hat{D}(h) = \mu_D(h) + \sum_{i=1}^{n_\alpha} {\alpha}_{i} \phi_i (h) , \label{eq:fpca_drag}
\end{equation}
\begin{equation}
    \hat{V}_{CAS}(h) = \mu_V(h) + \sum_{j=1}^{n_\beta}{\beta}_{j} \psi_j (h), \label{eq:fpca_cas}
\end{equation}
\end{subequations}
where $\mu_D$ and $\mu_V$ are mean functions and $\boldsymbol{\alpha}\in\Re^{n_\alpha}$ and $\boldsymbol{\beta}\in\Re^{n_\beta}$ a set of model weights. The functions $\phi(\cdot)$ and $\psi(\cdot)$ are orthonormal basis functions, satisfying the conditions: 
\begin{equation}
    \int \phi_i(h) \phi_j(h) dh = \delta_{ij}\; \text{and} \;\int \psi_k(h) \psi_l(h) dh = \delta_{kl},
\end{equation}
with $i,j\in[1,n_\alpha]$ and $k,l\in[1,n_\beta]$. The basis functions are discrete and defined over a discretised grid of flight levels. This grid is defined as $\boldsymbol{g}=[h_i, h_i+\Delta h, \dots, h_f]$. The first point in the grid, $h_i$, corresponds to the equivalent geodetic altitude of \gls{fl} 150, while $\Delta h$ is an increment of 1 \gls{fl} expressed in metres. The final grid point, $h_f$, is selected according to the range of \gls{fl}s for which data is available. \ref{app:gappy_pca} details the procedure of selecting $h_f$. The highest value of $h_f$ chosen in this paper was 12154 m, equivalent to \gls{fl} 398. The justification for selecting $h_i$ equivalent to \gls{fl} 150 is that UK airspace is heavily procedural below that level to handle arrivals and departures to and from a number of major airports. Focussing on the upper airspace allows a model that is more generalisable to be developed, whereas modelling individual standard arrival routes (STARs) would likely require a unique model be developed to account for speed limits and lateral manoeuvres that are specific to each STAR \cite{nats_fir}. 

\gls{fpca} follows similar principles to Principal Component Analysis (\gls{pca}). However, while \gls{pca} determines the eigenvalues and eigenvectors of a covariance matrix, estimated from samples, in \gls{fpca} a set of eigenvalues and eigenfunctions is sought for a covariance function. For more details on the mathematical background, the interested reader is referred to the canonical reference book of Ramsay and Silverman \cite{ramsay2013functional}. 

Trajectory data consists of a set of $n_b$ radar blips, denoted $\mathcal{X}=[\boldsymbol{x}^{(1)}, \dots, \boldsymbol{x}^{(n_b)}]$. Each radar blip, $\boldsymbol{x}\in\Re^5$ consists of quantities in the Mode S radar return that are relevant to descent in the \gls{bada} model, together with the inferred aircraft drag from \gls{bada}, i.e.:
\begin{enumerate}
    \item Geodetic altitude, $h$
    \item \gls{rocd}, $\frac{dh}{dt}$
    \item \gls{ias}, $V_{IAS}$
    \item Mach number, $M$
    \item Inferred drag, $\hat{D}$
\end{enumerate}

As has been discussed, \gls{ias}, denoted $V_{IAS}$, is a noisy measurement of \gls{cas}. Henceforth we assume $V_{IAS}\approx V_{CAS}$. We denote the dataset that contains data from $n_t$ trajectories as $\mathcal{T}=[\mathcal{X}^{(1)}, \dots, \mathcal{X}^{(n_t)}]$. Performing an \gls{fpca} analysis on the data in $\mathcal{T}$ requires every trajectory to be linearly interpolated onto the discretised grid of \gls{fl}s. This creates a challenge for trajectories that do not span the full range of \gls{fl}s. Trajectories for which this is the case are referred to in this paper as `gappy'. Two possible solutions are to either filter the trajectories in $\mathcal{T}$ such that only those that completely span the \gls{fl} range are used in the \gls{fpca} analysis or to restrict the studied \gls{fl} range such that the \gls{fl}s are common to all trajectories in $\mathcal{T}$. Both of these approaches place limitations on the amount of data that is available for model training in $\mathcal{T}$. 

To address this limitation, this paper proposes an  algorithm to impute missing portions of trajectories such that all trajectories in $\mathcal{T}$ span the studied \gls{fl} range. The algorithm was inspired by those of Everson and Sirovich \cite{everson1994karhunen} and Belda-Lois and Sánchez-Sánchez \cite{belda2015new} for \gls{fpca}. Details of the algorithm and its implementation are found in \ref{app:gappy_pca}. An important part of the procedure is determining the portion of the variance in the data which can be explained by the \gls{fpca} expansion, which effectively determines the truncation to $n_\alpha$ and $n_\beta$. Following a sweep over explained variance, a value of 80\% was chosen for this paper. More details on the results of this sweep may be found in \ref{app:exp_var}. 

\subsection{Probabilistic modelling of expansion weights}\label{sec:prob_models}
Computing the \gls{fpca} expansions for aircraft \gls{cas} and drag in descent yields a set of model weights, $\mathcal{W}=[\boldsymbol{w}^{(1)},\cdots,\boldsymbol{w}^{(n_t)}]$, where $\boldsymbol{w}^{(k)}\in\Re^{n_\alpha+n_\beta}$ contains the model weights computed for the $k$-th trajectory. These model weights form a low-dimensional latent space in which a probabilistic model can be trained to learn the joint distribution of $\mathcal{W}$. The probabilistic model can then be sampled to generate new drag and \gls{cas} functions to parameterise \gls{bada}, defining a new trajectory. 

Initial exploratory work in Pepper and Thomas \cite{pepper2023learning} fitted a Gaussian distribution to a set of \gls{fpca} coefficients for thrust. However, additional complexity has been introduced in this paper by jointly modelling \gls{cas} and drag. It is expected that a Gaussian distribution is too simple to capture the complex joint distribution. This paper explores a number of possible models for learning the joint distribution of $\mathcal{W}$: 
\begin{itemize}
    \item A Gaussian distribution, as in Pepper and Thomas \cite{pepper2023learning}.
    \item A Gaussian Mixture Model (\gls{gmm}), using the Bayesian Information Criterion (\gls{bic}) to select the number of uni-modal Gaussians in the mixture.
    \item A Normalizing Flow (\gls{nf}) model, a machine learning technique in which a series of neural networks form a parametrised transformation of a random variable \cite{papamakarios2021normalizing}.
\end{itemize}

The Gaussian model and \gls{gmm} were implemented in Python using the \texttt{sklearn} package \cite{scikit-learn}, while the normalizing flow was implemented using the \texttt{normflows} package \cite{Stimper2023}. Further details on the implementation of these probabilistic models, including the procedure used to determine the number of layers and neurons in the \gls{nf}s, are provided in \ref{app:num_params}.

\subsection{Ensuring physical plausibility}\label{sec:plaus}
Finally, it should be noted that the various probabilistic models considered are all unbounded. However, there are some combinations of model weights that will yield implausible trajectories, for example by no longer achieving the minimum \gls{rocd} mandated in UK airspace. Imposing plausibility requirements within the latent space is challenging due to the non-linear mapping between model weights and the corresponding trajectory when the \gls{bada} equations are solved. Instead, a set of bounds based on the data are computed for the drag and $V_{CAS}$ profiles. Sampled functions which violate these bounds are discarded and the model resampled. Heuristic bounds were computed from the data, chosen to be 95\% of the minimum and 105\% of the maximum value observed in the training data at that altitude.

\section{Data Preparation} \label{sec:data}

The data for this study comes from Mode S radar surveillance data collected between July to September 2019 in the London flight information region (FIR, \cite{nats_fir}), covering airspace over England and Wales. The data is classified by aircraft type and cleaned to select relevant descending portions of trajectories and remove any aircraft that return primary radar only, for example from General Aviation (GA) civil aircraft, military aircraft, and drones. 

Thirteen illustrative aircraft types are selected for this study, including aircraft with a range of aircraft mass, engine type, and \gls{icao} wake turbulence category (\cite{ClassJ}). The aircraft also vary in terms of how frequently they appear in UK airspace. The preliminary study in Pepper and Thomas \cite{pepper2023learning} only analysed aircraft types that frequently occurred in the dataset. This paper includes infrequently observed aircraft in order to study whether the performance of the model degrades significantly as training data in $\mathcal{T}$ and $\mathcal{W}$ becomes scarce. Properties of the studied aircraft are listed in Table~\ref{tab:actypes}. Aircraft engine types are either jet engines or turboprop engines, with jet engine aircraft far more prevalent in the dataset. The altitude range studied in this paper is above the altitude at which piston aircraft tend to operate, hence there are no piston engine types. The table also illustrates the maximum \gls{fl} observed in the dataset for each aircraft type and the equivalent geodetic altitude. Jet engined aircraft typically fly at much higher altitudes than turboprop aircraft. A range of \gls{icao} wake classes are represented in the data, including: heavy (H), light (L), medium (M), and super (J). 

\begin{table}%
\centering
\caption{Properties of the 13 studied aircraft. Turbo = turboprop.}
\label{tab:actypes}
\begin{tabular}{|c|c|c|c|c|c|c|}
\hline
Aircraft & $n_t$ & Wake turb.  & Mass,& Engine & Max. \gls{fl} & Max. $h$, \\
&& cat.&tonnes&&&m\\ \hline 
B738 & \num{41011} & M & \num{65}  & Jet &  \num{377} & \num{11490} \\
A320 & \num{28723} & M & \num{64}  & Jet &  \num{358} & \num{10910} \\
A319 & \num{20968} & M & \num{60}  & Jet &  \num{358} & \num{10910} \\
DH8D & \num{10821} & M & \num{26}  & Turbo &  \num{247} & \num{7529} \\
E190 & \num{6046} & M & \num{43}  & Jet &  \num{368} & \num{11220} \\
B772 & \num{2517} & H & \num{209}  & Jet &  \num{388} & \num{11830} \\
A388 & \num{1419} & J & \num{482}  & Jet &  \num{397} & \num{12100} \\
C56X & \num{1089} & M & \num{8}  & Jet &  \num{386} & \num{11770} \\
E170 & \num{1057} & M & \num{32}  & Jet &  \num{349} & \num{10640} \\
AT76 & \num{872} & M & \num{20}  & Turbo &  \num{188} & \num{5730} \\
PC12 & \num{701} & L & \num{4}  & Turbo &   \num{257} & \num{7833} \\
F2TH & 585%
& M & 16 & Jet & 398 & \num{12130}\\
BE20 & 257 & L & 5  & Turbo &  277 & \num{8443} \\
\hline
\end{tabular}
\end{table}

Prior to model fitting, the data was cleaned to select the free descent portions of aircraft trajectories. Firstly, those radar blips satisfying $\frac{dh}{dt}\leq -500 \text{ft/min}$ were retained. This is because the minimum descent rate within UK airspace is 500 ft/min. This selects descent trajectories once the free descent has been reached, excluding portions at the top and bottom of descent where the aircraft is either initiating the descent or levelling out. Additionally, trajectories are cleaned to remove portions of constant \gls{rocd} due to managed descent mode.

Aircraft descent profiles can be heavily affected by clearances issued by \gls{atc}. The goal of this paper is to develop a probabilistic model for aircraft in descent that are flying without ``level-by", climb rate, speed or procedural constraints.

Across the 13 aircraft types, 116,066 trajectories were retained. The aircraft type with the most available data is the B738 with $n_t$=41,011 while the least common aircraft was the BE20 with $n_t=${257}. Data was interpolated to a regular grid of flight levels such that \gls{fpca} could be performed. Given that the altitude range at which each aircraft type operates varies, a unique discretised grid of geodetic altitudes was used for each from \gls{fl} 150 or $h$={4,572} m to the aircraft specific maximum altitude in Table~\ref{tab:actypes}. An 80:20 train:test split was performed on the dataset before the \gls{fpca} was performed to generate the set of weights, $\mathcal{W}$, for every aircraft type. A separate probabilistic model was trained to learn the joint density of these weights for each aircraft type. As described in the previous section, a range of potential model types were explored for this step. The effectiveness of these models at representing the distribution of the held out test dataset is discussed in the next section. 

\section{Results} \label{sec:results} 
In this section the performance of the various methods for modelling the joint distribution of $\mathcal{W}$ was assessed through analysis of the generated trajectories. As was noted in the introduction, \gls{TP} methods proposed in the literature are typically deterministic, with performance evaluated using quantities such as the mean absolute error (\gls{mae}) and root mean squared error (see, e.g. \cite{wang:hal-01652041, de2013machine}). Such metrics are still relevant to the models developed in this paper. However, what is lacking in the literature is a set of benchmarks on which to assess the performance of probabilistic trajectory simulators. This paper suggests a number of metrics for quantifying the similarity of generated trajectories to held out real-world data that have domain-specific relevance to \gls{atc}.

Key findings were that the probabilistic model improved the mean \gls{cas} and drag predictions for all aircraft when compared with \gls{bada}. This leads to improved modelling of the descent rate and therefore estimates of the time to bottom of descent, with the proposed method providing at least a 7-fold improvement in the \gls{mae} (7.22 for the \gls{nf}). The \gls{nf} model produced distributions of trajectories that best matched the test data compared to the other probabilistic models. While in general trajectories generated by the proposed method appear realistic, sharp changes in \gls{rocd} that are observed in some turboprop trajectories are smoothed by the probabilistic model.

\subsection{Metrics for assessing probabilistic trajectory generators}\label{sec:metrics}
The quality of a probabilistic trajectory model should be assessed on two criteria:

\begin{itemize}
    \item How well the model captures the mean of the test trajectories.
    \item How well the synthetically generated sample trajectories reflect the distribution of the test trajectories.
\end{itemize}

To assess the first criterion in a manner that is relevant to \gls{atc} we compute the \gls{mae} for the time to bottom of descent. This is a relevant quantity for \gls{atc} as \gls{TP} methods are used within operational systems to (conservatively) predict future level occupancy to inform \gls{atco} decision making. Probability distributions of $V_{CAS}$ and \gls{rocd} are used to assess how well generated distributions match test data. $V_{CAS}$ is a relevant quantity to analyse as it acts as a proxy for the location of bottom of descent, which cannot be assessed explicitly without a model for wind conditions and the lateral portion of an aircraft's trajectory (which are not the subject of this paper). \gls{rocd} was analysed in distribution because, as a first derivative of the trajectory, it demands higher fidelity to achieve probabilistic equivalence compared to time to descend and $V_{CAS}$. To further probe the performance of the probabilistic model, statistical distances between the data and generated trajectories were computed above and below the calculated using \gls{bada} for each aircraft type. 

Probability Density Functions (\gls{pdf}s) can be computed using kernel density estimation for these quantities from generated trajectories and compared to the held out test dataset. There are a wealth of measures of statistical distance between probability distributions in the literature. In this paper the Kolmogorov-Smirnov (\gls{ks}) distance and Wasserstein are used. These metric distamces complement each other as the \gls{ks} distance quantifies the maximum discrepancy between the \gls{cdf}s of the distributions whilst the Wasserstein distance better quantifies the cost of transforming one distribution to the other.%

The next subsection displays generated trajectories for three aircraft types and compares these to the held out test data with the subsequent results computed on these generated samples. 

\subsection{Generating synthetic aircraft descents}
\begin{figure}
    \centering
    \includegraphics[width=\textwidth]{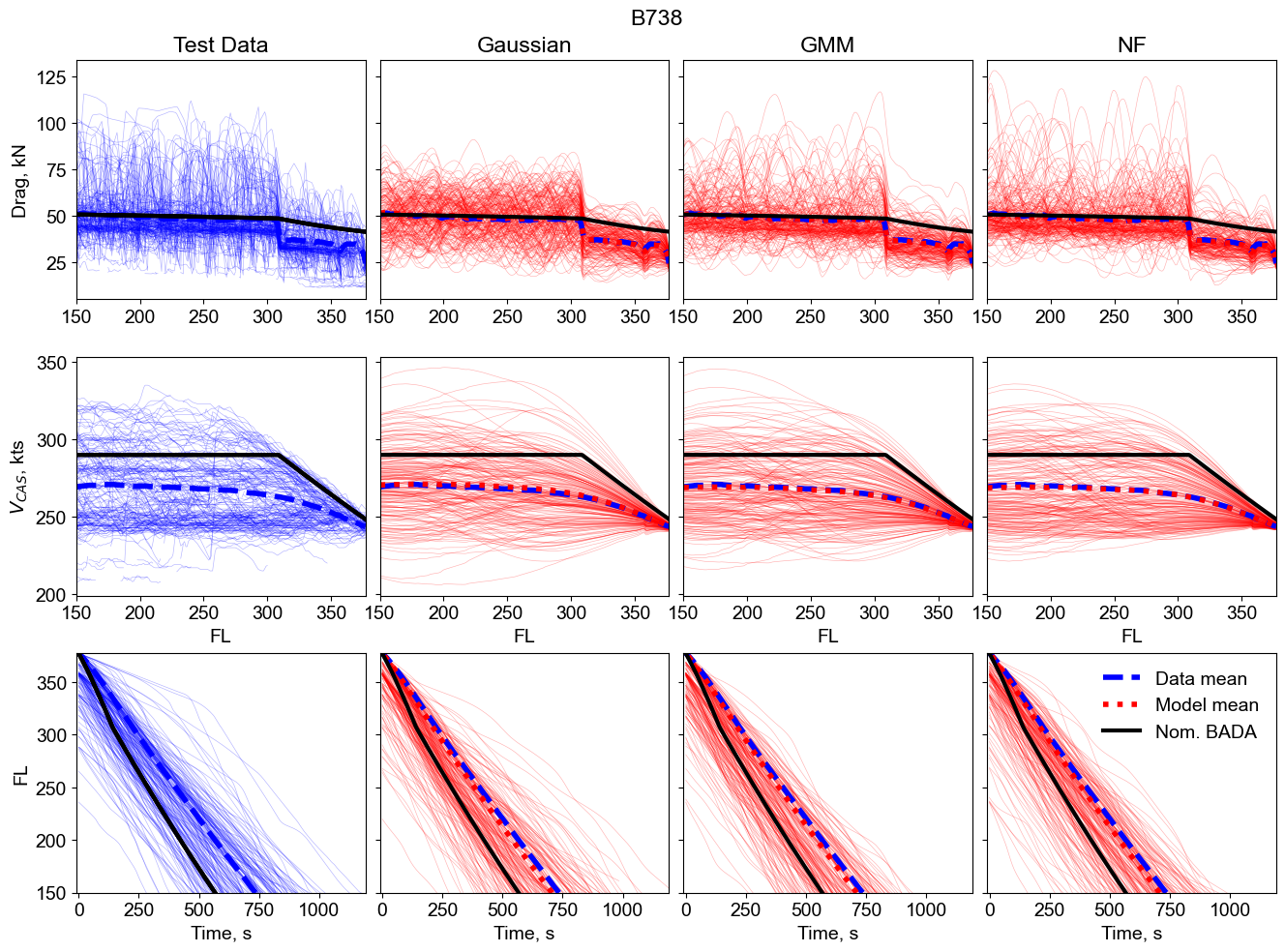}
    \caption{B738: held out test dataset (left column) and generated drag (top row), speed (middle row) and trajectories (bottom row) for the Gaussian, \gls{gmm}, and \gls{nf} based models.}
    \label{fig:B738_data}
\end{figure}

\begin{figure}
    \centering
    \includegraphics[width=\textwidth]{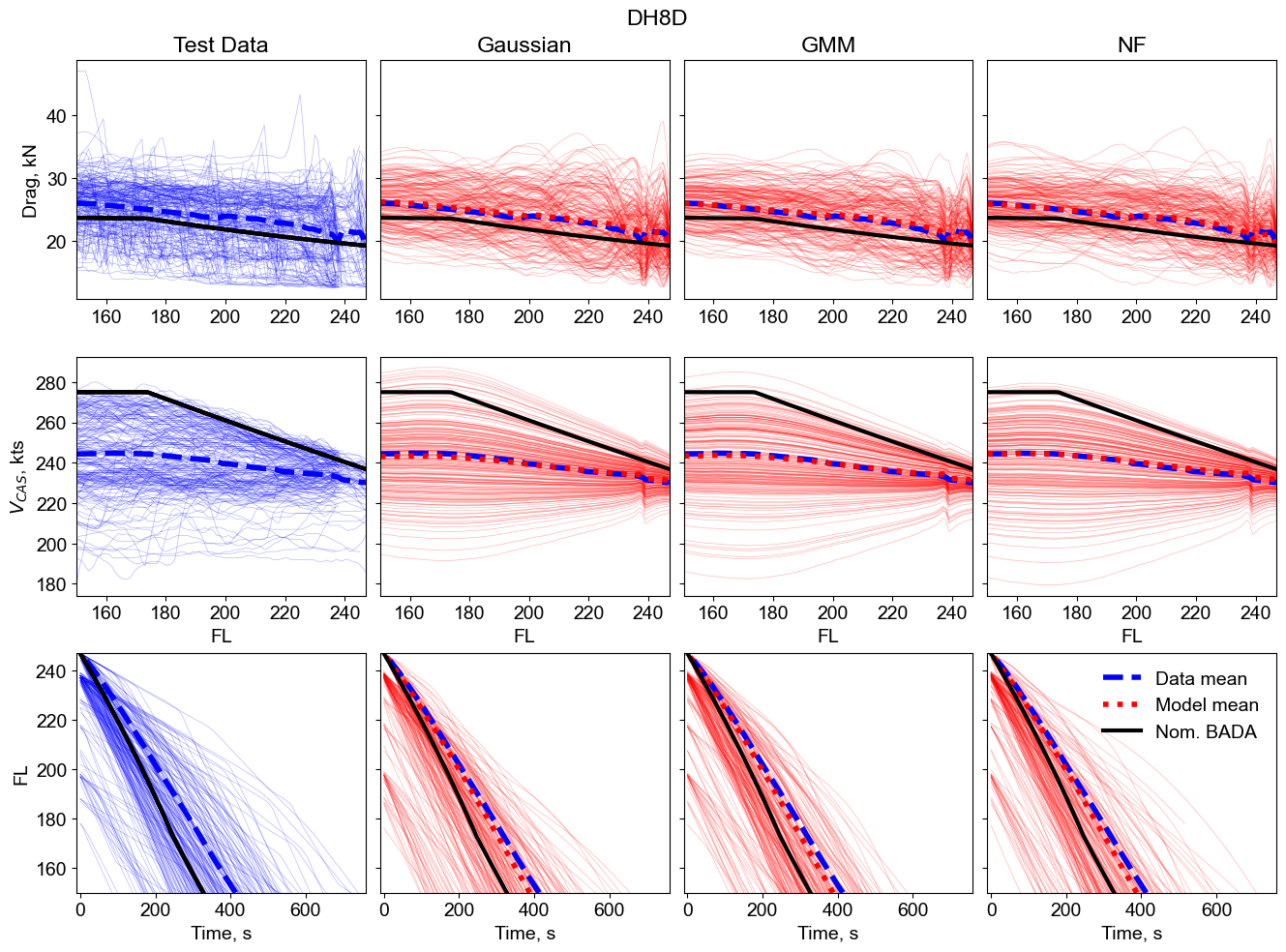}
    \caption{DH8D: held out test dataset (left column) and generated drag (top row), speed (middle row) and trajectories (bottom row) for the Gaussian, \gls{gmm}, and \gls{nf} based models.}
    \label{fig:DH8D_data}
\end{figure}

\begin{figure}
    \centering
    \includegraphics[width=0.8\textwidth]{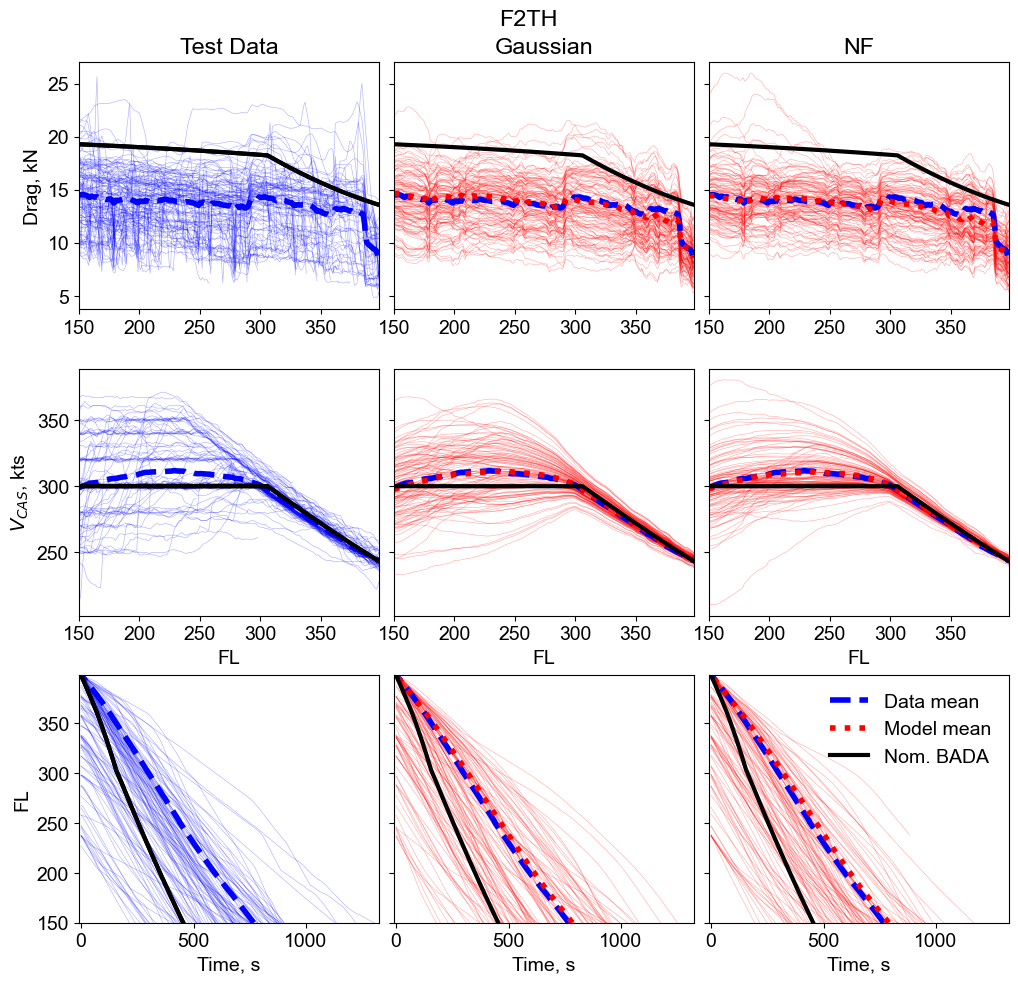}
    \caption{F2TH: held out test dataset (left column) and generated drag (top row), speed (middle row) and trajectories (bottom row) for the Gaussian and \gls{nf} based models.}
    \label{fig:F2TH_data}
\end{figure}

10,000 samples of the \gls{fpca} weights were generated from each probabilistic model and used to parameterise descents in \gls{bada}. As has been discussed in Section~\ref{sec:drag_cas}, the training data is gappy hence there are a range of potential initial \gls{fl}s in the training data. For the purposes of assessing model performance, the initial \gls{fl} of each generated sample was drawn from the distribution of starting altitudes in the test dataset. Using \eqref{eq:fpca}, these samples of the weights yield a set of samples of the drag and \gls{cas} functions that were used to parameterise \gls{bada}. Solving \eqref{eq:bada_rocd} using these drag and \gls{cas} functions generates a set of sampled descent trajectories. To illustrate the quality of the sampled trajectories, Figures~\ref{fig:B738_data}, \ref{fig:DH8D_data} and \ref{fig:F2TH_data} display the sampled functions and trajectories for the B738, DH8D, and F2TH aircraft types respectively. In each figure, the left-hand column of sub-figures illustrate the held out test dataset compared to the equivalent functions using nominal \gls{bada} (solid black line). The dashed blue line indicates the mean of the test data. The other columns of the sub-figures display the samples generated by the probabilistic model, together with the mean of the generated samples (dotted red line). These aircraft types were plotted as examples of a very frequently occurring jet aircraft (B738), a frequently occurring turboprop aircraft (DH8D), and an infrequently occurring jet aircraft with markedly different performance characteristics and wake class to the B738 (F2TH). 

\FloatBarrier

Some observations from visual inspection of Figures~\ref{fig:B738_data}, \ref{fig:DH8D_data} and \ref{fig:F2TH_data}: 

\begin{itemize}
    \item \textbf{Improved modelling of mean {\gls{cas}}}. The probabilistic models all capture the mean \gls{cas} of the test data well. This corrects noticeable misspecification in the \gls{bada} \gls{cas} speeds in Figures~\ref{fig:B738_data} and \ref{fig:DH8D_data}. This is to be expected as the parameters in the \gls{bada} model are calibrated globally, rather than for specific airspaces and time periods. The \gls{cas}-Mach transition altitude for both the B738 and DH8D is reflected in some of the data, but not all. For both aircraft the \gls{bada} \gls{cas} profile is seen in the data but it does not reflect the mean profile. \gls{bada} does capture the mean \gls{cas} profile well for the F2TH as showing in Figure~\ref{fig:F2TH_data}, although the probabalisitc model still improved on the mean prediction and captures variation in the \gls{cas} profile. Generally, for the studied aircraft, \gls{bada} overpredicts the aircraft speed.

    \item \textbf{Improved modelling of mean drag}. For all models and aircraft the mean prediction of the drag, which is computed from the samples, is close to the mean of the test data. \gls{bada} tends to overestimate the drag, although it underestimates for DH8D. In all the studied aircraft the combined \gls{cas} and drag misspecification leads to an underestimate of the total time to descend, or overestimate of the descent rate.
    
    \item \textbf{Generated trajectories appear realistic for the B738 and F2TH}. From visual inspection, generated trajectories for the B738 and the F2TH look realistic when compared with the test data. A number of DH8D trajectories are piecewise linear, with aircraft flying at constant \gls{rocd} for the initial part of the descent, before switching to a second constant \gls{rocd} that results in a steeper descent. The generated trajectories in Figure~\ref{fig:DH8D_data} are more curved as there are relatively few \gls{fpca} modes required to achieve 80\% explained variance cut-off for this aircraft type (see \ref{app:exp_var}). The number of retained modes is insufficient to capture this behaviour.
    
    \item \textbf{Behaviour in tails}. Some behaviour in the tails of the distribution of trajectories in the test dataset does not seem to have been captured. For instance, in Figure~\ref{fig:B738_data} several B738 trajectories initially descend relatively slowly, at a near constant \gls{rocd}, before descending at a similar rate to other trajectories. This behaviour is not captured in the probabilistic models. 
    
    \item \textbf{Smoothing of {\gls{cas}} profiles}. Due to the smoothing effect of the fPCA method, some \gls{cas} behaviour is not captured accurately. This effect is particularly pronounced in the speed profiles for the F2TH. Many speed profiles in Figure~\ref{fig:F2TH_data} are piecewise linear, with a knee at \gls{fl} 300, that is not captured in the generated samples. The samples from the NF appear the most realistic.
    
\end{itemize}
Section~\ref{sec:plaus} detailed the procedure by which sampled drag and \gls{cas} functions might be rejected if they exceeded a set of bounds defined over the grid $\boldsymbol{g}$. The median resample rate for most models was approximately 7\%. Higher resample rates were required for aircraft types with smaller training sets. For instance the BE20 had a resample rate of close to 9\% for the \gls{nf}, while the resample rate for aircraft types with larger training datasets, such as the B738, A320 and A319, were closer to 4-5\%. The DH8D had a high resample rate of 15\% for all models, likely due to the difficulty in representing turboprop descents using the \gls{fpca} representation. Resampling did not appreciably slow the speed of trajectory generation. The inference time for the Gaussian model was on the order of 4e-4~s for a MacBook Pro with an Apple M2 Pro chip. Inference times for the \gls{nf}s are longer, on the order of 2e-3~s. Solving the \gls{bada} equations using the sampled drag and \gls{cas} functions is a fast process, with the entire time to generate a trajectory on the order of 5e-3~s. In the following subsections, metrics are computed for the \gls{atc} relevant quantities discussed in \ref{sec:metrics}. 

\subsection{Time to bottom of descent}
Time to bottom of descent is an important quantity in \gls{atc}. Maintaining vertical separation between aircraft is a key task of controllers. Therefore, predictions of level occupancy with time in operational systems are important tools in supporting the decision making of controllers. To make the error comparable across trajectories, the time to bottom of descent was computed for the subset of trajectories in the test set that spanned the grid $\boldsymbol{g}$ (as defined in Section~\ref{sec:drag_cas}).

\begin{figure}
    \begin{subfigure}{\linewidth}
    \centering
    \includegraphics[width=1\linewidth]{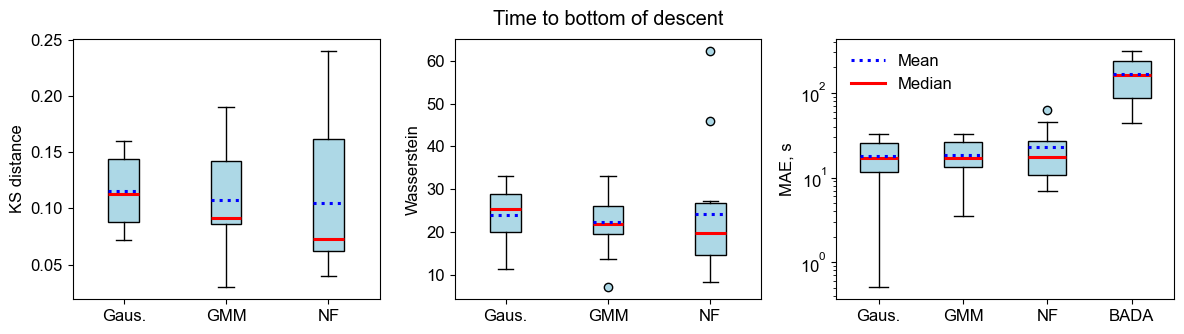}
    \caption{}%
    \label{fig:time_box}
    \end{subfigure}
    \begin{subfigure}{\linewidth}
    \centering
    \includegraphics[width=1\linewidth]{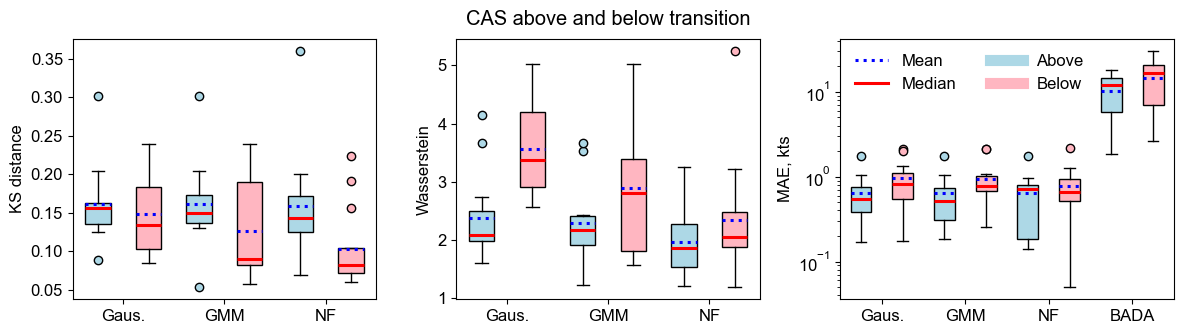}
    \caption{}%
    \label{fig:cas_box}
    \end{subfigure} 
    \begin{subfigure}{\linewidth}
    \centering
    \includegraphics[width=1\linewidth]{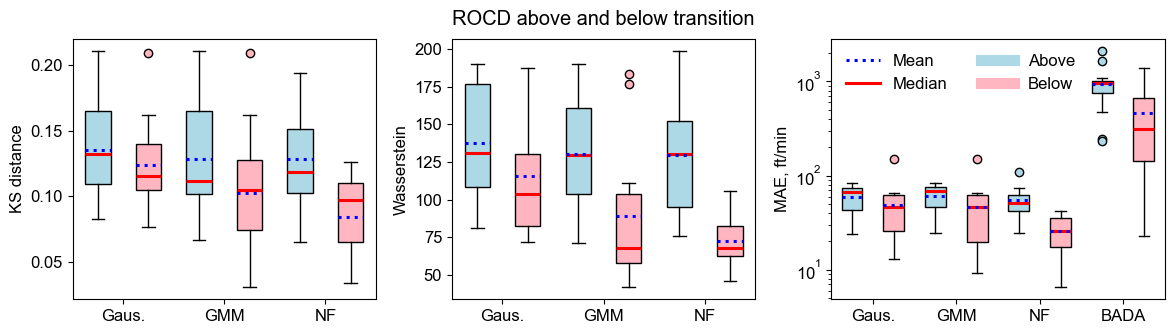}
    \caption{}%
    \label{fig:rocd_box}
    \end{subfigure}
    \caption{Boxplots showing the results of the three statistical distances on the time to bottom of descent (a), \gls{cas} (b), and \gls{rocd} (c) with \gls{ks} distance (left), Wasserstein distance (middle), and \gls{mae} (right).}
    \label{fig:boxplot}
\end{figure}

Figure~\ref{fig:time_box} displays box plots of the time to bottom of descent for each of the probabilistic models. Each box contains the performance of each model across all 13 aircraft types. The left and centre subplots illustrate boxplots of the \gls{ks} and Wasserstein distance between the distribution of the time to bottom of descent from the probabilistic models compared to the test dataset. The median error of the \gls{nf} is significantly lower than the other methods, although the performance is more variable between aircraft types than the other methods which is likely due to over-fitting on smaller datasets. The right subplot in Figure~\ref{fig:time_box} indicates the \gls{mae} in predicting the time to bottom of descent. All probabilistic models exceed the performance of nominal \gls{bada} in predicting this quantity by a factor of 8.7, averaged across the three models\footnote{{The mean \gls{mae} of time to bottom of descent is 166 s for \gls{bada}, compared to 18 s, 16 s, and 23 s for the Gaussian, \gls{gmm} and \gls{nf} models respectively.}}. The \gls{gmm} had the best overall \gls{mae} for this quantity, while the \gls{nf} had the lowest statistical distance to the test dataset of the methods.  

\begin{figure}[ht!]
    \begin{subfigure}{0.5\linewidth}
    \centering
    \includegraphics[width=1\linewidth]{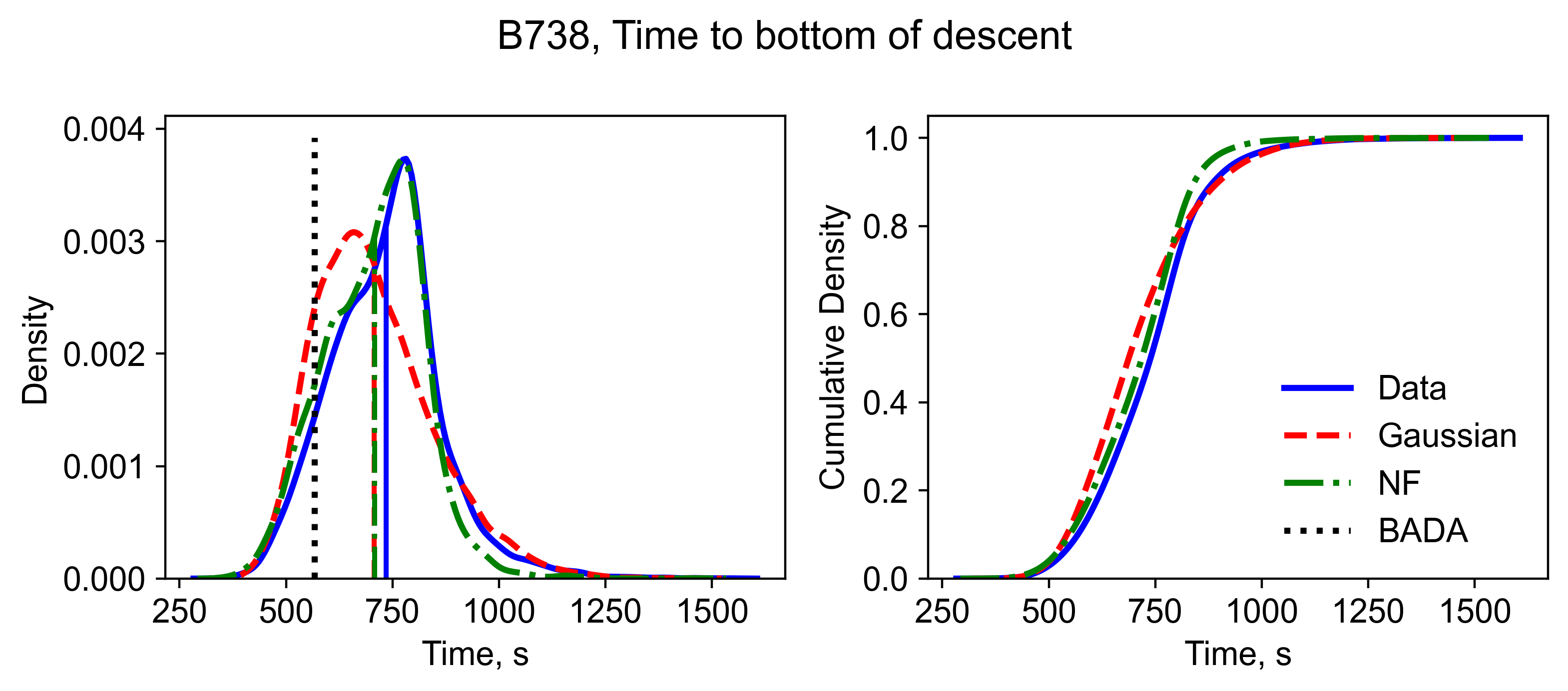}
    \caption{}
    \label{fig:cuts_time_b738}
    \end{subfigure}
    \begin{subfigure}{0.5\linewidth}
    \includegraphics[width=1\linewidth]{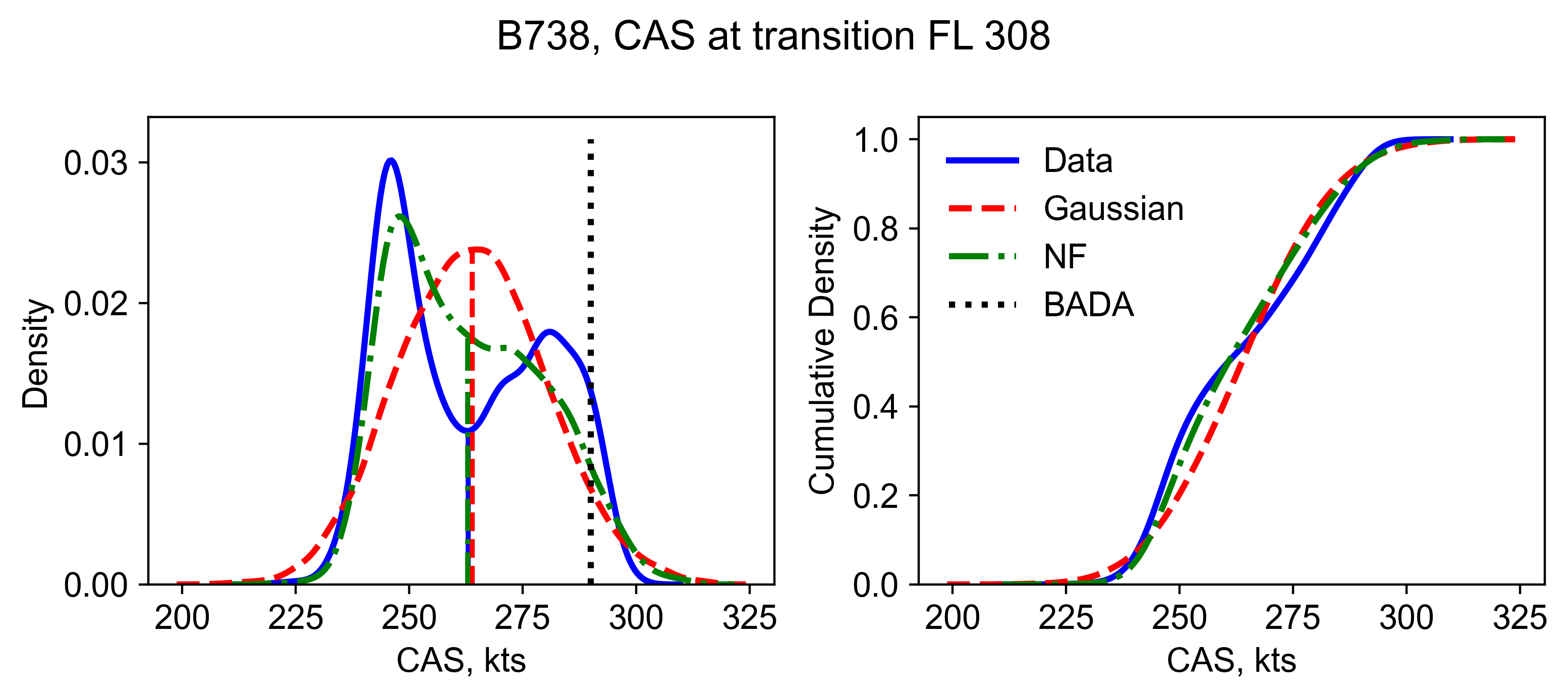}
    \caption{}
    \label{fig:cuts_cas_b738}
    \end{subfigure}
    \begin{subfigure}{0.5\linewidth}
    \includegraphics[width=1\linewidth]{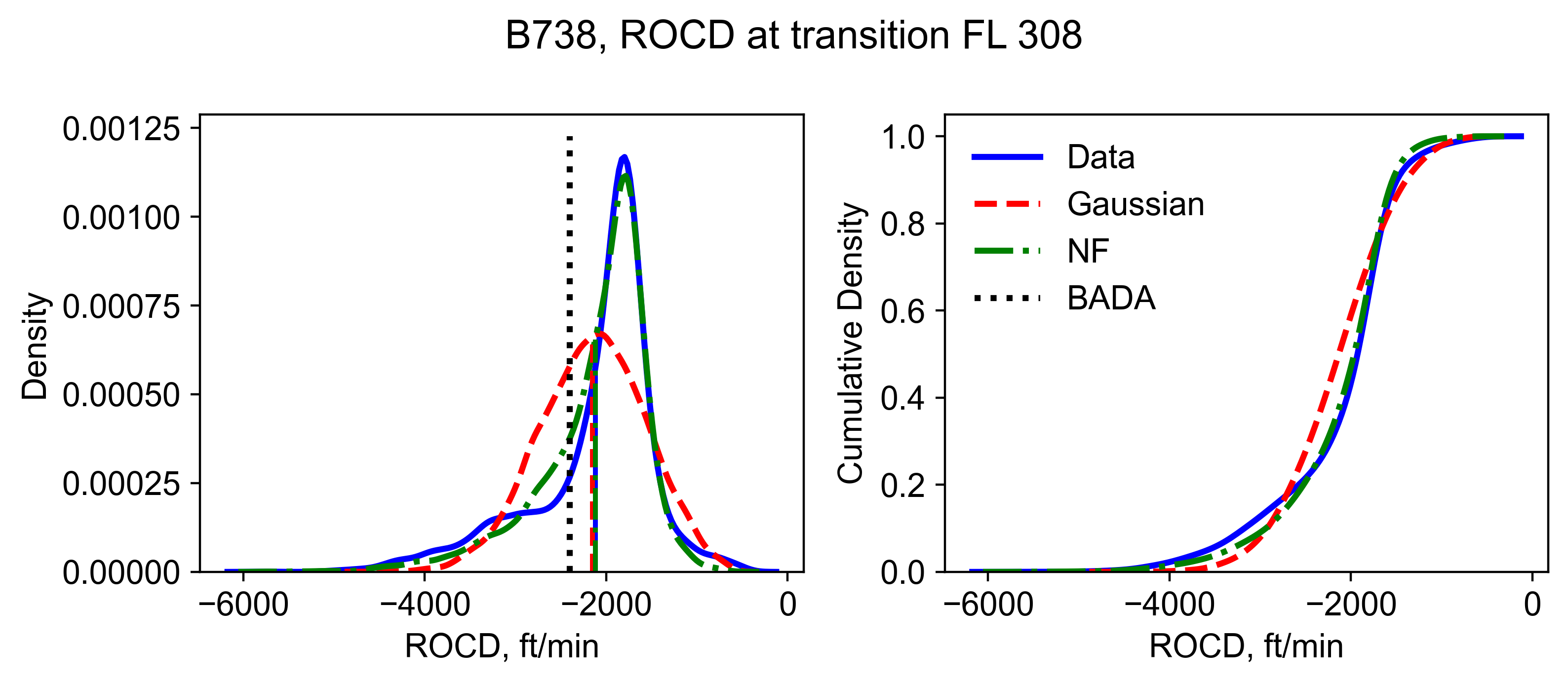}
    \caption{}
    \label{fig:cuts_rocd_b738}
    \end{subfigure}%
    \caption{Distribution of time to bottom of descent, \gls{cas} and \gls{rocd} at the \gls{cas}-Mach transition point for B738 for data and synthetic trajectories with the Gaussian and \gls{nf} model. \gls{bada} prediction and means shown with the vertical lines.}
    \label{fig:cuts_b738}
\end{figure}

\begin{figure}[ht!]
    \begin{subfigure}{0.5\linewidth}
    \centering
    \includegraphics[width=1\linewidth]{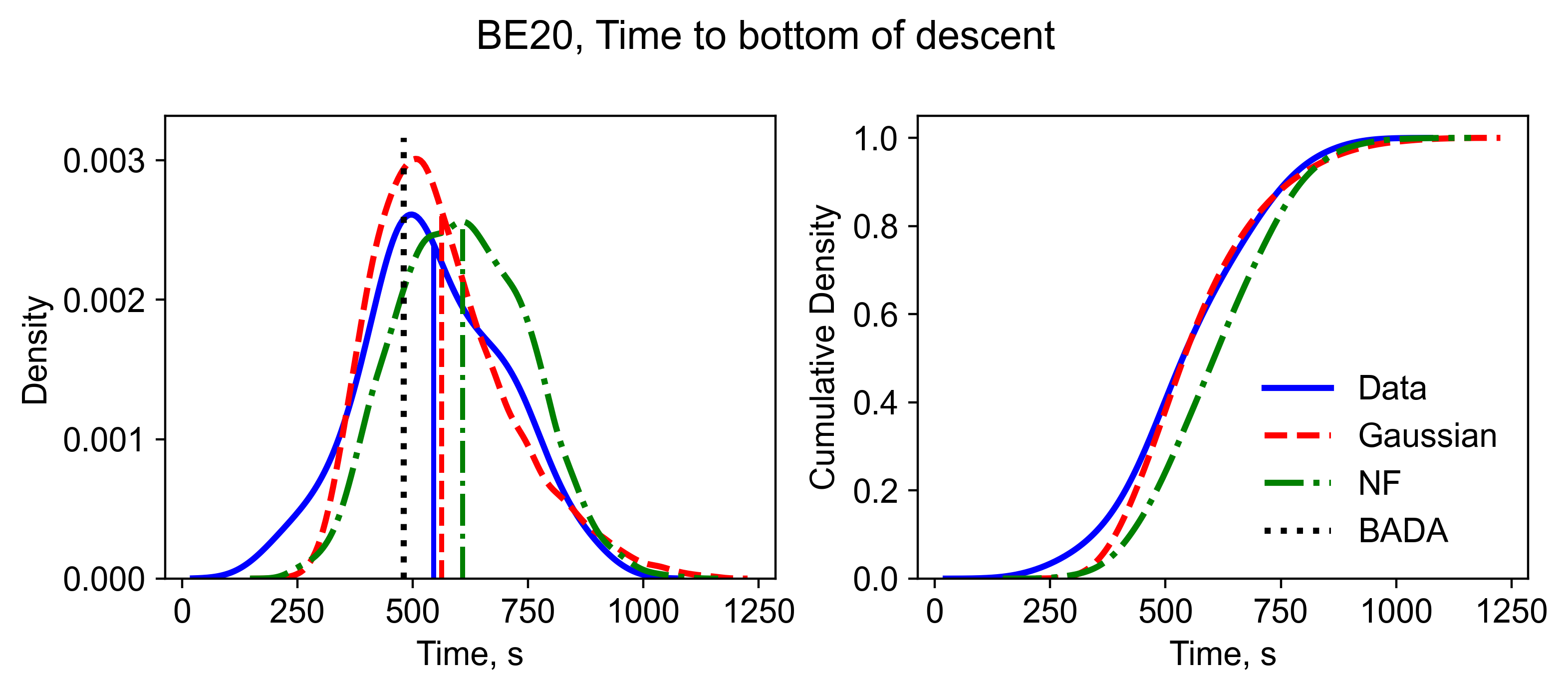}
    \caption{}
    \label{fig:cuts_time_be20}
    \end{subfigure}
    \begin{subfigure}{0.5\linewidth}
    \centering
    \includegraphics[width=1\linewidth]{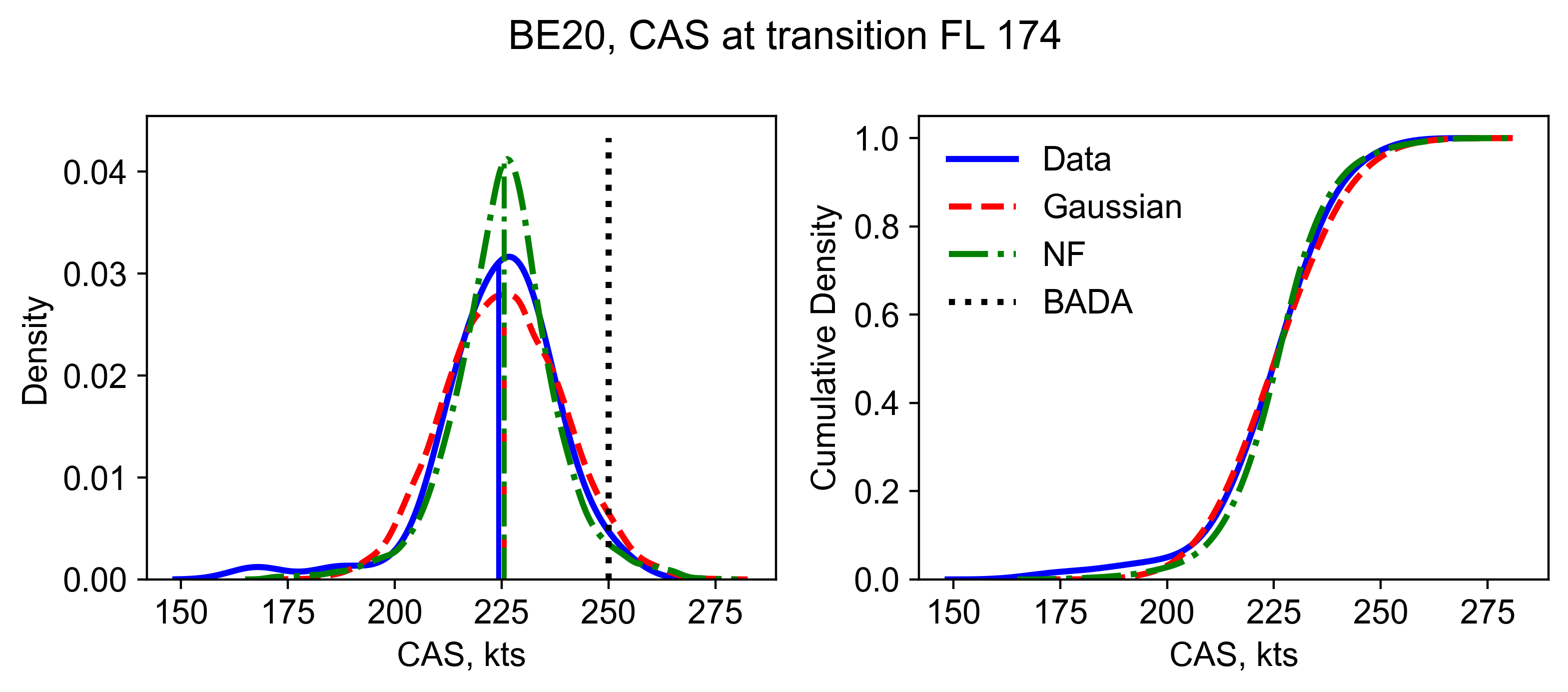}
    \caption{}
    \label{fig:cuts_cas_be20}
    \end{subfigure}
\begin{subfigure}{0.5\linewidth}
    \includegraphics[width=1\linewidth]{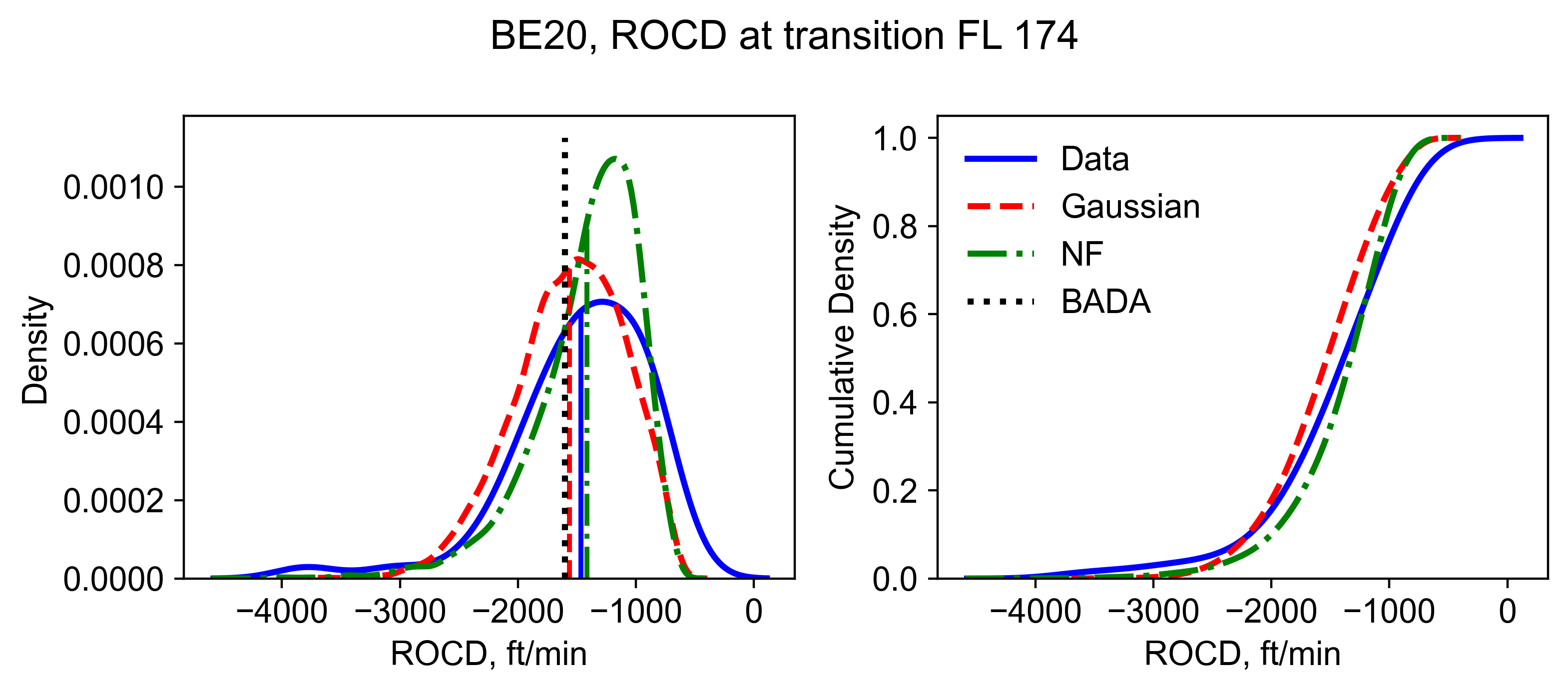}
    \caption{}
    \label{fig:cuts_rocd_be20}
    \end{subfigure}   
    
    \caption{Distribution of time to end of descent, and \gls{cas} and \gls{rocd} at the \gls{cas}-Mach transition point for BE20 for data and synthetic trajectories with the Gaussian and \gls{nf} model. \gls{bada} prediction and means shown with the vertical lines.}
    \label{fig:cuts_be20}
\end{figure}

Figure \ref{fig:cuts_time_b738} displays the \gls{pdf} and empirical cumulative distribution function (\gls{cdf}) for the time to bottom of descent for the B738. Figure~\ref{fig:cuts_time_be20} is an equivalent plot for the BE20. \gls{pdf}s and \gls{cdf}s of the test data are indicated by solid lines and are compared to those of the trajectories generated by the Gaussian (dashed lines) and \gls{nf} (dot-dash lines). The dashed vertical line represents the \gls{bada} prediction. For both aircraft types the mean time to descend is under-predicted by \gls{bada}. Figure \ref{fig:cuts_time_b738} shows that the distribution of time to descend of the \gls{nf} is closer to that of the test dataset, although the modelling of the right-hand tail is actually superior in the Gaussian. Similarly, in Figure~\ref{fig:cuts_time_be20}, the distribution of time to descend for the Gaussian generated trajectories is closer to the test dataset than the \gls{nf}. The BE20 dataset is much smaller than that of the B738 ($n_t=257$ compared to 41,011) suggesting that the relative performance of the methods is to some extent conditioned on the size of the dataset used for training and testing. Full results for the metrics are tabulated in \ref{app:table_results}.

\subsection{CAS in descent above and below transition point}\label{sec:res_cas}
As has been discussed, \gls{cas} in descent may be thought of as a proxy for assessing predictions of the location of bottom of descent, in lieu of a model of the aircraft route and a wind model to convert the \gls{cas} into ground speed. The various probabilistic models were assessed according to the statistical distance between the $V_{CAS}$ in the generated trajectories and that of the test dataset. Unlike the time to bottom of descent, the error in the distribution of \gls{cas} was meaningful for all trajectories and the entire test dataset could be used to compute this metric. 

Consideration of imbalance in the trajectory dataset was made when computing this. If raw radar blips are used to compute a distribution of $V_{CAS}$, trajectories that descend more slowly account for more of the dataset as they contain more radar blips than a trajectory that descends more quickly over the same altitude range. The following procedure was used to compute statistical distance between the test dataset and generated trajectories: 

\begin{enumerate}
    \item Interpolate $V_{CAS}$ to the levels in $\boldsymbol{g}$ for each trajectory in the test dataset. 
    \item Interpolate $V_{CAS}$ to the levels in $\boldsymbol{g}$ for each trajectory in the generated dataset. 
    \item Compute the statistical distance (\gls{ks} or Wasserstein distance or \gls{mae}) at each level in $\boldsymbol{g}$.
    \item Report the mean statistical distance across those levels in grid $\boldsymbol{g}$ that are above the transition point, with a separate mean taken over levels below the transition altitude.
\end{enumerate}
This approach allows for the performance of the models above and below the transition point to be probed, while also rebalancing the dataset to account for variations in number of radar blips between trajectories. Figure~\ref{fig:cas_box} shows boxplots for the statistical distance between distributions of $V_{CAS}$. Pairs of boxes are plotted for each probabilistic model, with the left-hand box corresponding to the average statistical distance above transition (light blue) and the right-hand box the statistical distances below transition (pink). It should be noted that the FL range for the AT76 are fully located below transition and these aircraft therefore do not provide a value above transition.

The right hand panel of Figure~\ref{fig:cas_box} indicates significant improvement (a factor of 18.2\footnote{The mean \gls{mae} across aircraft types is 10.4 kts for \gls{bada} above transition, compared to 0.65 kts, 0.42 kts, and 0.65 kts for the Gaussian, \gls{gmm} and \gls{nf} models respectively. Below transition the mean \gls{bada} \gls{mae} is 14.7 kts, compared to 0.96 kts, 0.81 kts, and 0.77 kts for the Gaussian, \gls{gmm}, and \gls{nf} models.}) in the \gls{mae} of $V_{CAS}$ for the probabilistic models compared to the nominal \gls{bada} model. The mean \gls{mae} of both above and below \gls{cas} predictions being less than 1~kt for all probabilistic models. Boxes for the \gls{ks} distance are consistent both across probabilistic models and also between portions of the dataset above and below the transition altitude. On the other hand, when computing statistical distance with the Wasserstein distance, the performance in the probabilistic models degrades below transition. The \gls{nf} is clearly superior to the other probabilistic models above transition, and comparable to the \gls{gmm} below.

Some context for why the performance can be dependent on the measure of statistical distance is provided by Figures~\ref{fig:cuts_cas_b738}, which display \gls{pdf}s and \gls{cdf}s of $V_{CAS}$ at the transition altitude for the B738. From visual inspection of the \gls{pdf}s, there is clearly significant discrepancy between both the \gls{nf} and Gaussian \gls{pdf}s and that of the data. However, because the supremum distance between the \gls{cdf} of the data and that of the \gls{nf} is relatively small, the KS distance is 0.08 for the \gls{nf} compared to 0.15 for the Gaussian. On the other hand, the discrepancy in performance as measured by the Wasserstein distance is less dramatic (3.85 for the Gaussian and 2.18 for the \gls{nf}) as the Wassterstein distance effectively integrates the discrepancies between \gls{pdf}s. 

Figure \ref{fig:cuts_cas_be20} displays the \gls{pdf}s and \gls{cdf}s for the BE20 $V_{CAS}$ at the transition altitude. From inspection, the matching between the models and data appears much better for the BE20. Unlike the equivalent plot for the B738 in Figure~\ref{fig:cuts_cas_b738}, the distribution of \gls{cas} in Figure~\ref{fig:cuts_cas_be20} is uni-modal, rather than bi-modal and is well approximated by the Gaussian model. In both Figures~\ref{fig:cuts_cas_b738} and \ref{fig:cuts_cas_be20} the mean $V_{CAS}$ in the data is appreciably lower than the nominal \gls{cas} in the \gls{bada} model\footnote{{For the B738, the \gls{mae} for \gls{cas} at transition for \gls{bada} is 27.0 kts, compared to 0.8~kts and 0.15 kts for the Gaussian and \gls{nf} models respectively. For the BE20 the \gls{mae} for \gls{bada} was 25.8 kts for \gls{bada} and 1.3 kts and 1.4 kts for the Gaussian and \gls{nf}.}}.

\subsection{ROCD computed at discrete FLs}\label{sec:rocd_test}

Finally, statistical distances between distributions of \gls{rocd} were computed above and below the transition point using the procedure outlined in Section~\ref{sec:res_cas}. Figure~\ref{fig:rocd_box} displays boxplots of the statistical distance across aircraft types, with separate boxes for altitudes above and below the transition altitude. The rationale for computing statistical distances for \gls{rocd} was that, as a first derivative of each trajectory, it would require a high-fidelity model in order to achieve probabilistic equivalence. It is a meaningful quantity for all trajectories, even those that are gappy. In fact \gls{ks} distances in Figure~\ref{fig:rocd_box} are comparable to those for $V_{CAS}$ in Figure~\ref{fig:cas_box} (the Wasserstein distance cannot be compared across quantities of interest with different units). Again the mean error of the \gls{nf} was lower than that of the Gaussian and \gls{gmm}s, but all models achieved at least a 10-fold mean improvement in \gls{mae} compared to \gls{bada}. Interestingly, model performance does not degrade below the transition altitude as it did for $V_{CAS}$. The \gls{gmm} and \gls{nf} have a lower Wasserstein distance below transition than above it.

Figures~\ref{fig:cuts_rocd_b738} and~\ref{fig:cuts_rocd_be20} show the \gls{pdf} and \gls{cdf}s for \gls{rocd} at the transition altitude for the B738 and BE20 respectively. In both cases, the Gaussian and \gls{nf} model captures the data mean well\footnote{{The \gls{mae} for \gls{rocd} at transition for \gls{bada} is 280 ft/min for the B738, compared to 30 ft/min and 0 ft/min for the Gaussian and \gls{nf} models. The \gls{bada} \gls{mae} is 134 ft/min for the BE20, compared to 98 ft/min and 51 ft/min for the Gaussian and \gls{nf}.}}. However, the \gls{nf} captures the distribution of the B738 test data almost exactly, unlike in the BE20 case\footnote{{The \gls{ks} distance for \gls{rocd} at transition for the \gls{nf} is 0.0506 for the B738 and 0.141 for the BE20.}}. Again, this is likely due to the relative scarcity of data in the BE20 case. 

\FloatBarrier

\section{Conclusions} \label{concs}
This paper presents a method for simulating plausible aircraft descents that match real world data. The hybrid model used a models terms in a physics-based model to ensure generated trajectories are physically plausible. Drag and \gls{cas} profiles are generated with a probabilistic model that is trained in the latent space formed by the weights of \gls{fpca} representations of drag and \gls{cas}. The models were trained on a dataset of 116,066 trajectories from Mode S returns of aircraft in UK airspace over a 3 month period in 2019. Thirteen aircraft types were investigated with varying propulsion, mass, and performance characteristics. 

Various options for the form of the probabilistic model were evaluated by comparing generated trajectories to a held out dataset of real world data. The realism of the generated distributions of trajectories was assessed by analysing probability distributions of three quantities  relevant to either \gls{atc} or accurate trajectory modelling. Simpler models for generating trajectories in the latent space such as Gaussian Mixture Models were compared to Normalizing Flows for this task. Regardless of model choice, the mean of the proposed method was significantly closer to the held out training dataset than the \gls{bada} model. In terms of estimating time to reach bottom of descent, the proposed methods improve the \gls{mae} error 10-fold compared with \gls{bada}, averaged across the studied aircraft types and all three models. Estimates of this quantity were improved for all aircraft types. The mean predictions of \gls{cas} and \gls{rocd}, above and below the transition point, also improved by at least an order of 10 compared to the \gls{bada} baseline.

Analysis of the statistical distance between 10,000 generated trajectories and the held out test datasets suggest a reasonable level of agreement. However, this is difficult to quantify without additional probabilistic methods for generating descents to act as a baseline. The normalizing flow models gave a closer match to the test dataset for the majority of aircraft types, compared to the simpler models evaluated. Factors such as the modality of the target distribution and the quantity of available training data influenced the relative performance of the models from one aircraft type to the next.

This work has presented a method for generating realistic profiles of descending aircraft in an airspace that is dominated by jet aircraft. As such the model has been primarily designed to model unconstrained descents of jet aircraft and has been validated by assessing how well real-world trajectories are matched when viewed in distribution. The logic required to handle descents under speed instructions or model specific known flight management system descent modes is beyond the scope of this paper but is additional fidelity that could be added to the presented method in the future. 

\section*{Acknowledgement}
The authors would like to thank \textbf{Ben Carvell, Martin Layton, and Richard Everson} for their helpful comments. The work described in this paper is primarily funded by the grant “EP/V056522/1: Advancing Probabilistic Machine Learning to Deliver Safer, More Efficient and Predictable Air Traffic Control” (aka Project Bluebird), an EPSRC Prosperity Partnership between NATS, The Alan Turing Institute, and the University of Exeter.

\bibliographystyle{dcu}
\bibliography{main.bib}

@article{pepper2023learning,
author = {Pepper, Nick and Thomas, Marc},
title = {Learning Generative Models for Climbing Aircraft from Radar Data},
journal = {Journal of Aerospace Information Systems},
volume = {21},
number = {6},
pages = {474-481},
year = {2024},
doi = {10.2514/1.I011359},
URL = {\url{https://doi.org/10.2514/1.I011359}}
}

@article{pepper2023probabilistic,
  title={{A probabilistic model for aircraft in climb using monotonic functional Gaussian process emulators}},
  author={Pepper, Nick and Thomas, Marc and De Ath, George and Olivier, Enrico and Cannon, Richard and Everson, Richard and Dodwell, Tim},
  journal={Proceedings of the Royal Society A},
  volume={479},
  number={2271},
  pages={20220607},
  year={2023},
  publisher={The Royal Society},
  url={\url{https://doi.org/10.1098/rspa.2022.0607}}
}

@article{papamakarios2021normalizing,
  title={Normalizing flows for probabilistic modeling and inference},
  author={Papamakarios, George and Nalisnick, Eric and Rezende, Danilo Jimenez and Mohamed, Shakir and Lakshminarayanan, Balaji},
  journal={The Journal of Machine Learning Research},
  volume={22},
  number={1},
  pages={2617--2680},
  year={2021},
  publisher={JMLRORG},
  url={\url{{http://jmlr.org/papers/v22/19-1028.html}}}
}

@article{nuic2010bada,
  title={BADA: An advanced aircraft performance model for present and future ATM systems},
  author={Nuic, Angela and Poles, Damir and Mouillet, Vincent},
  journal={International journal of adaptive control and signal processing},
  volume={24},
  number={10},
  pages={850--866},
  year={2010},
  publisher={Wiley Online Library},
    url={\url{https://doi.org/10.1002/acs.1176}}
}

@article{nuic2010user,
  title={User manual for the Base of Aircraft Data (BADA) revision 3.10},
  author={Nuic, Angela},
  journal={Atmosphere},
  volume={2010},
  pages={001},
  year={2010},
  url={\url{http://maartenuijtdehaag.com/bada310-user-manual.pdf}}
}

@article{bastas2020data,
  title={Data driven aircraft trajectory prediction with deep imitation learning},
  author={Bastas, Alevizos and Kravaris, Theocharis and Vouros, George A},
  journal={arXiv preprint arXiv:2005.07960},
  year={2020},
    url={\url{https://doi.org/10.48550/arXiv.2005.07960}}
}

@article{pang2021data,
  title={Data-driven trajectory prediction with weather uncertainties: A Bayesian deep learning approach},
  author={Pang, Yutian and Zhao, Xinyu and Yan, Hao and Liu, Yongming},
  journal={Transportation Research Part C: Emerging Technologies},
  volume={130},
  pages={103326},
  year={2021},
  publisher={Elsevier},
  url={\url{https://doi.org/10.1016/j.trc.2021.103326}}
}

@article{thipphavong2013adaptive,
  title={Adaptive algorithm to improve trajectory prediction accuracy of climbing aircraft},
  author={Thipphavong, David P and Schultz, Charles A and Lee, Alan G and Chan, Steven H},
  journal={Journal of Guidance, Control, and Dynamics},
  volume={36},
  number={1},
  pages={15--24},
  year={2013},
  publisher={American Institute of Aeronautics and Astronautics},
  url={\url{https://doi.org/10.2514/1.58508}}
}

@inproceedings{de2013machine,
  title={A machine learning approach to trajectory prediction},
  author={De Leege, Arjen and van Paassen, Marinus and Mulder, Max},
  booktitle={AIAA Guidance, Navigation, and Control (GNC) Conference},
  pages={4782},
  year={2013}, 
    url={\url{https://doi.org/10.2514/6.2013-4782}}
}

@article{ma2020hybrid,
  title={{A hybrid CNN-LSTM model for aircraft 4D trajectory prediction}},
  author={Ma, Lan and Tian, Shan},
  journal={IEEE access},
  volume={8},
  pages={134668--134680},
  year={2020},
  publisher={IEEE},
    url={\url{https://api.semanticscholar.org/CorpusID:220886893}}
}

@article{shi20204,
  title={{4-D flight trajectory prediction with constrained LSTM network}},
  author={Shi, Zhiyuan and Xu, Min and Pan, Quan},
  journal={IEEE transactions on intelligent transportation systems},
  volume={22},
  number={11},
  pages={7242--7255},
  year={2020},
  publisher={IEEE},
  url={\url{https://ieeexplore.ieee.org/abstract/document/9136843}}
}

@article{lymperopoulos2010sequential,
  title={{Sequential Monte Carlo methods for multi-aircraft trajectory prediction in air traffic management}},
  author={Lymperopoulos, Ioannis and Lygeros, John},
  journal={International Journal of Adaptive Control and Signal Processing},
  volume={24},
  number={10},
  pages={830--849},
  year={2010},
  publisher={Wiley Online Library},
    url={\url{https://doi.org/10.1002/acs.1174}}
}

@article{alligier2013learning,
  title={Learning the aircraft mass and thrust to improve the ground-based trajectory prediction of climbing flights},
  author={Alligier, Richard and Gianazza, David and Durand, Nicolas},
  journal={Transportation Research Part C: Emerging Technologies},
  volume={36},
  pages={45--60},
  year={2013},
  publisher={Elsevier},
    url={\url{https://doi.org/10.1016/j.trc.2013.08.006}}
}

@inproceedings{nicol2013functional,
  TITLE = {{Functional principal component analysis of aircraft trajectories}},
  AUTHOR = {Nicol, Florence},
  URL = {\url{https://enac.hal.science/hal-00867957}},
  BOOKTITLE = {{ISIATM 2013, 2nd International Conference on Interdisciplinary Science for Innovative Air Traffic Management}},
  ADDRESS = {Toulouse, France},
  YEAR = {2013},
  MONTH = Jul,
  HAL_ID = {hal-00867957},
  HAL_VERSION = {v1},
}

@article{krauth2023deep,
  title={Deep generative modelling of aircraft trajectories in terminal maneuvering areas},
  author={Krauth, Timoth{\'e} and Lafage, Adrien and Morio, J{\'e}r{\^o}me and Olive, Xavier and Waltert, Manuel},
  journal={Machine Learning with Applications},
  volume={11},
  pages={100446},
  year={2023},
  publisher={Elsevier},
    url={\url{https://api.semanticscholar.org/CorpusID:254439547}}
}

@inproceedings{pang2020conditional,
  title={{Conditional generative adversarial networks (CGAN) for aircraft trajectory prediction considering weather effects}},
  author={Pang, Yutian and Liu, Yongming},
  booktitle={AIAA Scitech 2020 Forum},
  pages={1853},
  year={2020},
  url={\url{https://doi.org/10.2514/6.2020-1853}}
}

@article{tran2022aircraft,
  title={Aircraft trajectory prediction with enriched intent using encoder-decoder architecture},
  author={Tran, Phu N and Nguyen, Hoang QV and Pham, Duc-Thinh and Alam, Sameer},
  journal={IEEE Access},
  volume={10},
  pages={17881--17896},
  year={2022},
  publisher={IEEE}, 
  url={\url{https://ieeexplore.ieee.org/document/9704880}}
}

@article{everson1994karhunen,
  title={Karhunen--Loeve procedure for gappy data},
  author={Everson, Richard and Sirovich, Lawrence},
  journal={JOSA A},
  volume={12},
  number={8},
  pages={1657--1664},
  year={1995},
  publisher={Optica Publishing Group},
    url={\url{https://doi.org/10.1364/JOSAA.12.001657}}
}

@inproceedings{belda2015new,
  title={A new methodology for Functional Principal Component Analysis from scarce data. Application to stroke rehabilitation},
  author={Belda-Lois, Juan-Manuel and S{\'a}nchez-S{\'a}nchez, M Luz},
  booktitle={2015 37th Annual International Conference of the IEEE Engineering in Medicine and Biology Society (EMBC)},
  pages={4602--4605},
  year={2015},
  organization={IEEE}, 
    url={\url{https://pubmed.ncbi.nlm.nih.gov/26737319/}}
}

@book{ramsay2013functional,
  title={Functional Data Analysis},
  author={Ramsay, J. and Silverman, B.W.},
  isbn={9781475771077},
  lccn={96054729},
  series={Springer Series in Statistics},
  url={\url{https://books.google.co.uk/books?id=fgLqBwAAQBAJ}},
  year={2013},
  publisher={Springer New York}
}

@online{nats_fir,
  author  = {NATS},
  title   = {Introduction to airspace},
  year    = {2023},
  url     = {\url{https://www.nats.aero/ae-home/introduction-to-airspace/}},
  urldate = {17-05-2023}
}

@online{ClassJ,
  title   = {{Regulatory Article (RA) 3277: wake turbulence}},
  author  = {{Ministry of Defence (UK) and Military Aviation Authority }},
  year    = {2015},
  url= {\url{https://www.gov.uk/government/publications/regulatory-article-ra-3277-wake-turbulence}}
}

@article{orlando1989mode,
  title={{The mode S beacon radar system}},
  author={Orlando, Vincent A},
  journal={The Lincoln Laboratory Journal},
  volume={2},
  number={3},
  pages={345--362},
  year={1989},
  publisher={Citeseer},
    url={\url{https://www.ll.mit.edu/sites/default/files/publication/doc/mode-s-beacon-radar-system-orlando-ja-6373.pdf}}
}

@article{Stimper2023, 
  author = {Vincent Stimper and David Liu and Andrew Campbell and Vincent Berenz and Lukas Ryll and Bernhard Schölkopf and José Miguel Hernández-Lobato}, 
  title = {{normflows: A PyTorch Package for Normalizing Flows}}, 
  journal = {Journal of Open Source Software}, 
  volume = {8},
  number = {86}, 
  pages = {5361}, 
  publisher = {The Open Journal}, 
  doi = {10.21105/joss.05361}, 
  url = {\url{https://doi.org/10.21105/joss.05361}}, 
  year = {2023}
}

@article{scikit-learn,
  title={{Scikit-learn: Machine Learning in Python}},
  author={Pedregosa, F. and Varoquaux, G. and Gramfort, A. and Michel, V.
          and Thirion, B. and Grisel, O. and Blondel, M. and Prettenhofer, P.
          and Weiss, R. and Dubourg, V. and Vanderplas, J. and Passos, A. and
          Cournapeau, D. and Brucher, M. and Perrot, M. and Duchesnay, E.},
  journal={Journal of Machine Learning Research},
  volume={12},
  pages={2825--2830},
  year={2011},
    url={\url{{http://jmlr.org/papers/v12/pedregosa11a.html}}}
}

@article{papamakarios2017masked,
  title={Masked autoregressive flow for density estimation},
  author={Papamakarios, George and Pavlakou, Theo and Murray, Iain},
  journal={Advances in neural information processing systems},
  volume={30},
  year={2017},
  url={\url{https://doi.org/10.48550/arXiv.1705.07057}}
}

@online{Mats_part1,
  title   = {{CAP 493: Manual of Air Traffic Services (MATS) Part 1}},
  author  = {{UK Civil Aviation Authority}},
  year    = {2023},
  url= {\url{https://www.caa.co.uk/our-work/publications/documents/content/cap-493/}}
}

@online{easr_2022,
  title   = {{2022 European Aviation Environmental Report}},
  author  = {{European Union Aviation Safety Agency (EASA)}},
  year    = {2022},
  url= {\url{https://www.easa.europa.eu/eco/eaer}}
}

@online{Sesar_sol,
  author  = {{SESAR Joint Undertaking}},
  title   = {{SESAR} solutions catalogue (Fourth edition)},
  year    = {2021},
  url     = {\url{https://www.sesarju.eu/activities-solutions}},
  urldate = {24-03-2023}
}

@online{cdo,
  title   = {{Continuous climb and descent operations}},
  author  = {{Eurocontrol}},
  year    = {2024},
  url= {\url{https://www.eurocontrol.int/concept/continuous-climb-and-descent-operations}}
}

@inproceedings{riihimaki2010gaussian,
  title={Gaussian processes with monotonicity information},
  author={Riihim{\"a}ki, Jaakko and Vehtari, Aki},
  booktitle={Proceedings of the thirteenth international conference on artificial intelligence and statistics},
  pages={645--652},
  year={2010},
  organization={JMLR Workshop and Conference Proceedings},
  url={\url{http://proceedings.mlr.press/v9/riihimaki10a.html}}
}

@article{barratt2018learning,
  title={Learning probabilistic trajectory models of aircraft in terminal airspace from position data},
  author={Barratt, Shane T and Kochenderfer, Mykel J and Boyd, Stephen P},
  journal={IEEE Transactions on Intelligent Transportation Systems},
  volume={20},
  number={9},
  pages={3536--3545},
  year={2018},
  publisher={IEEE},
  url={\url{https://ieeexplore.ieee.org/abstract/document/8551278}}
}

@article{pepper_sector,
  title={A Probabilistic Method for Sector-Specific 4d Aircraft Trajectory Prediction},
  author={ Pepper, Nick and De Ath, George and Carvell, Ben and Hodgkin, Amy and Dodwell, Tim and Thomas, Marc and Everson, Richard},
  journal={SSRN},
  year=2024,
  url={\url{https://ssrn.com/abstract=4984556}}
}

@inproceedings{wang:hal-01652041,
  TITLE = {{Short-term 4D Trajectory Prediction Using Machine Learning Methods}},
  AUTHOR = {Wang, Zhengyi and Liang, Man and Delahaye, Daniel},
  URL = {\url{https://enac.hal.science/hal-01652041}},
  BOOKTITLE = {{SID 2017, 7th SESAR Innovation Days}},
  ADDRESS = {Belgrade, Serbia},
  YEAR = {2017},
  MONTH = Nov,
  HAL_ID = {hal-01652041},
  HAL_VERSION = {v1},
}

@article{wu2022long,
  title={Long-term 4D trajectory prediction using generative adversarial networks},
  author={Wu, Xiping and Yang, Hongyu and Chen, Hu and Hu, Qinzhi and Hu, Haoliang},
  journal={Transportation Research Part C: Emerging Technologies},
  volume={136},
  pages={103554},
  year={2022},
  publisher={Elsevier},
  url={\url{https://doi.org/10.1016/j.trc.2022.103554}}
}

@online{sesar,
  author  = {{SESAR Joint Undertaking}},
  title   = {Delivering the Digital European Sky},
  year    = {2021},
  url     = {\url{https://www.sesarju.eu/}},
  urldate = {24-03-2023}
}

@article{willard2022integrating,
  title={Integrating scientific knowledge with machine learning for engineering and environmental systems},
  author={Willard, Jared and Jia, Xiaowei and Xu, Shaoming and Steinbach, Michael and Kumar, Vipin},
  journal={ACM Computing Surveys},
  volume={55},
  number={4},
  pages={1--37},
  year={2022},
  publisher={ACM New York, NY}
}

\appendix
\section{Algorithm for gappy fPCA} \label{app:gappy_pca}
This section outlines the main steps of the algorithm that was used to determine the \gls{fpca} basis when the trajectories in $\mathcal{T}$ are gappy, i.e. they do not cover the full geodetic altitude range in $\boldsymbol{g}$. Recall that $\boldsymbol{g}\in\Re^{n_g}$ is a uniformly spaced grid of geodetic altitudes between $h_i$ and $h_f$. As a preprocessing step, trajectories in $\mathcal{T}$ are interpolated onto $\boldsymbol{g}$ and collected in the (sparse) data matrix $F\in\Re^{n_t\times n_g}$. The gappy algorithm requires the covariance matrix, $C\in\Re^{n_g\times n_g}$, to be computed at each iteration. After the first iteration of the algorithm, $C$ can be computed using:
\begin{equation}
    C=\frac{1}{n_t-1} (F-\mu_F)^T (F-\mu_F), 
\end{equation}
where $\mu_F$ denotes the column-wise mean of $F$. However, for the first algorithm a sparse representation of $F$ is used. In this instance $C$ is approximated using the non-zero elements of $F$. In what follows, the gappy algorithm is explained in terms of determining the basis functions $\phi_i(h)$ and weights $\boldsymbol{\alpha}$ for the drag function (see \eqref{eq:fpca_drag} and \eqref{eq:fpca_cas}). The process can be repeated to determine the basis and weights of $V_{CAS}$. 

The main steps of the gappy \gls{fpca} algorithm are as follows:

\begin{enumerate}
    \item Compute covariance matrix, $C$.
    \item Determine the principal components of $C$ with an eigenvalue decomposition of the covariance matrix, i.e.
    \begin{equation}
    C\Phi=\boldsymbol{\lambda} C,
    \end{equation}
    where $\Phi$ is a matrix containing the eigenvectors and $\boldsymbol{\lambda}$ a vector of eigenvalues. The eigenvectors represent the modes in a discrete \gls{fpca} basis. 
    \item Weights associated with the $k$-th trajectory are determined through a sequential quadratic programming approach, with the $l$-th weight found through the optimisation procedure:
    \begin{equation}
        \hat{\alpha}_l^{(k)}=\underset{\boldsymbol{\alpha}^{(k)}}{\rm{arg\,min}} \; \sum_{i=1}^{n_g} \Big[F_{ki}-\mu_{F_{ki}}-\sum_{j=1}^l\alpha_j^{(k)}\Phi_{ij} \Big]^2,
    \end{equation}
    with the weight associated with each \gls{fpca} mode found in turn. 
    \item The drag function can be reconstructed using \eqref{eq:fpca_drag} together with the fitted weights, effectively imputing the missing values in the data. 
    \item The process is repeated using the reconstructed data until the average reconstructed data changes by less than 1\% every iteration for 10 iterations, up to a maximum of 50 iterations. 
\end{enumerate}

\begin{figure}
    \centering
    \includegraphics[width=0.5\linewidth]{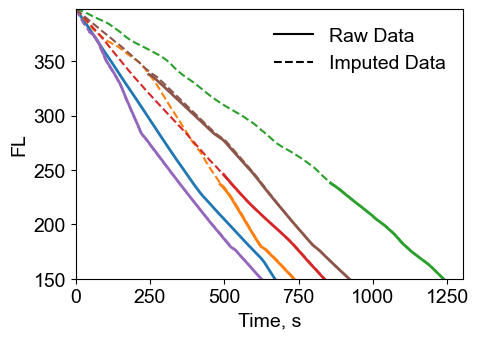}
    \caption{F2TH trajectories from gappy data plotted with their full reconstructed trajectories using all fPCA modes}
    \label{fig:fpca_working}
\end{figure}

When constructing $\boldsymbol{g}$, the highest geodetic altitude, $h_f$, is chosen as the greatest geodetic altitude present in at least 25\% of the trajectories in the dataset. Figure~\ref{fig:fpca_working} shows a sample of 5 gappy trajectories for the F2TH aircraft along with the full trajectories imputed using all of the \gls{fpca} modes.

\section{fPCA explained variance selection} \label{app:exp_var} 

The \gls{fpca} expansions in Section \ref{sec:drag_cas} (\eqref{eq:fpca_drag} and \eqref{eq:fpca_cas}) are sums over two sets of orthonormal bases, truncated to include $n_{\alpha}$ and $n_\beta$ orthonormal basis functions. This truncation is determined by the percentage of variance in the dataset that the \gls{fpca} basis can explain. In this work we select the $n_\alpha$ and $n_\beta$ that correspond to 80\% explained variance. 

To justify this choice a sweep over explained variance was performed using the training dataset described in Section~\ref{sec:data}. The effect of varying the number of components in the \gls{fpca} expansion was investigated by training a \gls{gmm} and testing it on the held out fold of test data. A 5-fold cross validation procedure was used to train and test models. With 5 folds and 13 aircraft types, this corresponded to training and testing 65 \gls{gmm}s for every choice of explained variance. 1,000 samples were generated by each model in order to compute the statistical distance to the held out fold. 

\begin{figure}
    \centering
    \includegraphics[width=\textwidth]{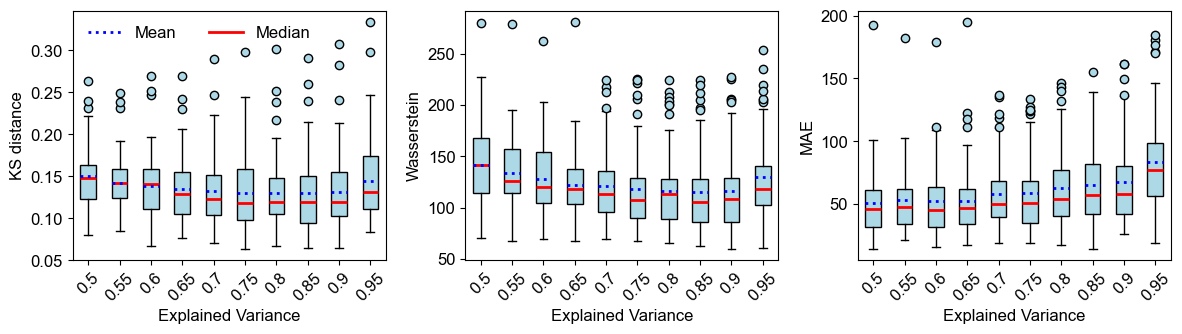}
    \caption{Boxplots showing a sweep of explained variance, with error computed between distributions of the \gls{rocd}.}
    \label{fig:exp_var_fpca}
\end{figure}

Figure \ref{fig:exp_var_fpca} is a boxplot displaying the results of the sweep over explained variance. The \gls{ks} distance, Wasserstein distance and \gls{mae} were used to compute error in the \gls{rocd}, as in Section~\ref{sec:rocd_test}. As the choice of explained variance is increased from 50\% (0.5 in Figure \ref{fig:exp_var_fpca}), the mean and median error drops in the \gls{ks} and Wasserstein distances, while the \gls{mae} stays relatively constant. Beyond an explained variance of 75-80\% an overfitting effect occurs and both the median Wasserstein distance and \gls{mae} increases with increasing explained variance. Based on this sweep an explained variance of 80\% was used in this paper. 

\section{Determining model sizes} \label{app:num_params}
Section~{\ref{sec:prob_models}} lists the various probabilistic models that were evaluated for learning the joint distribution of $\mathcal{W}$. As shown in Table~\ref{tab:actypes} the number of aircraft types in the dataset, $n_t$, varies significantly between aircraft types. Note that the actual number of aircraft in the training dataset is $n_{tr}=0.8n_t$ due to the train:test split, using $n_{tr}$ to represent the number of trajectories in the training set. We denote $n_w=n_\alpha+n_\beta$ as the dimensionality of the joint weight space and use the following procedures to determine a suitable number of parameters to use in each probabilistic model. 

\hspace{1pt} \\
\noindent \textbf{Gaussian model} \\
The mean and covariance function of a Gaussian distribution is fitted using the expectation maximization algorithm from the Python package \texttt{sklearn} \cite{scikit-learn}. The matrix is symmetric, therefore only the upper or lower triangle needs to be fitted, with the number of model parameters, $n_p$, determined by:
\begin{equation}
    n_p = n_c(n_c + 3)/2.
\end{equation}
It was found that over-parameterisation was not a concern for the Gaussian model as $n_p\ll n_{tr}$ for all aircraft types. 

\hspace{1pt} \\
\noindent \textbf{Gaussian mixture model} \\
The \gls{gmm} is more flexible than the Gaussian model as multiple Gaussian models can be included in the mixture, with $n_m$ denoting the number of individual Gaussian models in the mixture. The number of model parameters scales as: 
\begin{equation}
    n_p = n_{m}n_c(n_c + 3)/2 + (n_{m}-1),
\end{equation}
with $n_m$ and $n_c$. The second term on the right-hand side corresponds to the parameters used to weight the components of the mixture relative to one another. The Bayesian Information Criterion (\gls{bic}) was used to determine $n_m$, up to a maximum permissible value corresponding to $15 n_p\leq n_{tr}$, i.e. at least 15 trajectories were required per parameter in the \gls{gmm}. Again the \texttt{sklearn} Python package was used for implementation. Table~\ref{tab:gmm_comps} contains information on the number of components selected for the \gls{gmm}s based on the \gls{bic}. Note that in the case of 5 aircraft types $n_{tr}$ was such that $n_m$ was limited to 1, i.e. a uni-modal Gaussian distribution. In all but one case $n_m$ is equal to the maximum permissible value, suggesting a need for a higher fidelity model to capture the distribution. 

\begin{table}[htbp]%
\centering
\begin{tabular}{|c|c|c|c|c|c|c|}
\hline
Aircraft & $n_{tr}$ & $n_\alpha$ & $n_\beta$  & $n_c$ & Max. $n_m$  & $n_m$ from BIC \\ \hline
B738 & 32808 & 13 & 2 & 15  & 16 & 16  \\
A320 & 22978 & 10 & 2 & 12  & 16 & 16  \\
A319 & 16774 & 10 & 2 & 12  & 12 & 12  \\
DH8D & 8656 & 4 & 1 & 5  & 27 & 19  \\
E190 & 4836 & 6 & 2 & 8  & 7 & 7  \\
B772 & 2013 & 11 & 1 & 12  & 1 & 1  \\
A388 & 1135 & 11 & 2 & 13  & 1 & 1  \\
E170 & 845 & 5 & 2 & 7  & 1 & 1  \\
C56X & 871 & 4 & 2 & 6  & 2 & 2  \\
PC12 & 560 & 3 & 3 & 6  & 1 & 1  \\
AT76 & 697 & 2 & 2 & 4  & 3 & 3  \\
F2TH & 468 & 4 & 2 & 6  & 1 & 1  \\
BE20 & 205 & 3 & 2 & 5  & 1 & 1  \\
\hline
\end{tabular}
\caption{Number of GMM components, $n_m$, used for each aircraft type.}
\label{tab:gmm_comps}
\end{table}

\hspace{1pt} \\
\noindent \textbf{Normalizing Flow} \\
Normalizing flows use a neural network architecture to map samples from a normal distribution, denoted $p(\boldsymbol{x})$, to a transformed probability distribution $p(\boldsymbol{u|\boldsymbol{x}})$, with $\boldsymbol{x}, \boldsymbol{u}\in\Re^{n_c}$, through a series of invertible mappings. By training the network hyperparameters to minimise the Kullback-Leibler (\gls{kl}) divergence between $\mathcal{W}$ and samples from $p(\boldsymbol{u|\boldsymbol{x}})$, a \gls{nf} can be trained to emulate a complex distributions of real-world data \cite{papamakarios2021normalizing}. 

A \gls{nf} is structured as a series of neural networks, with each network referred to as a `flow'. Each flow has a single hidden layer and input and output layers with $n_c$ neurons. A coarse sweep over number of flows and numbers of neurons in the hidden layer was performed, with best results for \gls{nf}s with 5 flows and $2n_c$ neurons in the hidden layers. Masked Affine Flows are used, which employ Masked Autoencoder for Distribution Estimation units to dropout connections \cite{papamakarios2017masked}.   

\FloatBarrier

\section{Tabulated results} \label{app:table_results}
This section presents tabulated results for the metrics described in Section~\ref{sec:results}. Table~\ref{tab:stats_time} presents results for the time to bottom of descent for each of the investigated aircraft types. Statistical distances for \gls{cas} above and below transition are provided in Tables~\ref{tab:stats_ias_above_av} and~\ref{tab:stats_ias_below_av}. Finally, Tables~\ref{tab:stats_rocd_above_av}~and~\ref{tab:stats_rocd_below_av} tabulate the results for \gls{rocd} above and below the transition point. For each measure of statistical distance, the best performing probabilistic model for that aircraft type is circled, with dashed boxes indicating two or more models within 5\% of each other. As a deterministic model, \gls{bada} is only evaluated using the \gls{mae}. Dashes indicate that there was insufficient data to train \gls{gmm}s with $n_m>1$ using the procedure outlined in \ref{app:num_params}. 

\begin{table}[h!]\fontsize{9}{9.3}\selectfont
\caption{Time to bottom of descent for each aircraft type.}
\label{tab:stats_time}
\centering \begin{tabular}{|c|c|c|c|c|c|}
\hline
AC Type & Statistic & Gaussian & GMM & NF & BADA \\
 \hline
B738 & KS distance & 0.158 & 0.0858 & \ovalbox{0.0765} & \\
B738 & Wasserstein & 31.8 & \odashbox{24.3} & \odashbox{26.8} & \\
B738 & MAE, s & 28.3 & \odashbox{24.3} & \odashbox{26.8} & 168 \\\hline
A320 & KS distance & 0.11 & \odashbox{0.0882} & \odashbox{0.0918} & \\
A320 & Wasserstein & \ovalbox{22.5} & 26.0 & 27.2 & \\
A320 & MAE, s & \ovalbox{22.2} & 26.0 & 27.2 & 160 \\\hline
A319 & KS distance & 0.0723 & \ovalbox{0.0297} & 0.0401 & \\
A319 & Wasserstein & 12.9 & \ovalbox{7.07} & 8.33 & \\
A319 & MAE, s & \ovalbox{1.48} & 6.98 & 6.87 & 141 \\\hline
DH8D & KS distance & \ovalbox{0.160} & 0.190 & 0.180 & \\
DH8D & Wasserstein & \odashbox{25.3} & \odashbox{25.9} & \odashbox{25.6} & \\
DH8D & MAE, s & \odashbox{25.3} & \odashbox{25.9} & \odashbox{25.5} & 85.9 \\\hline
E190 & KS distance & 0.0778 & \odashbox{0.0665} & \odashbox{0.0655} & \\
E190 & Wasserstein & 26.4 & \odashbox{20.0} & \odashbox{19.8} & \\
E190 & MAE, s & 16.9 & \ovalbox{14.8} & 17.2 & 184 \\\hline
B772 & KS distance & 0.113 & - & \ovalbox{0.0614} & \\
B772 & Wasserstein & 19.6 & - & \ovalbox{10.5} & \\
B772 & MAE, s & 17.9 & - & \ovalbox{9.38} & 162 \\\hline
A388 & KS distance & 0.144 & - & \ovalbox{0.0704} & \\
A388 & Wasserstein & 32.2 & - & \ovalbox{17.0} & \\
A388 & MAE, s & 31.9 & - & \ovalbox{16.5} & 43.7 \\\hline
E170 & KS distance & \ovalbox{0.144} & - & 0.182 & \\
E170 & Wasserstein & \ovalbox{33.0} & - & 45.8 & \\
E170 & MAE, s & \ovalbox{33.0} & - & 45.8 & 245 \\\hline
C56X & KS distance & 0.0882 & 0.0785 & \ovalbox{0.0624} & \\
C56X & Wasserstein & 25.7 & 17.8 & \ovalbox{14.2} & \\
C56X & MAE, s & \ovalbox{0.504} & 3.49 & 10.7 & 236 \\\hline
PC12 & KS distance & 0.142 & - & \ovalbox{0.0622} & \\
PC12 & Wasserstein & 20.1 & - & \ovalbox{14.7} & \\
PC12 & MAE, s & 11.5 & - & \ovalbox{9.50} & 282 \\\hline
AT76 & KS distance & \ovalbox{0.118} & 0.136 & 0.162 & \\
AT76 & Wasserstein & \ovalbox{11.4} & 13.8 & 18.1 & \\
AT76 & MAE, s & \ovalbox{10.5} & 13.2 & 17.6 & 71.6 \\\hline
F2TH & KS distance & 0.0916 & - & \ovalbox{0.0731} & \\
F2TH & Wasserstein & 29.0 & - & \ovalbox{24.1} & \\
F2TH & MAE, s & \ovalbox{16.0} & - & 21.8 & 314 \\\hline
BE20 & KS distance & \ovalbox{0.0872} & - & 0.240 & \\
BE20 & Wasserstein & \ovalbox{21.8} & - & 62.3 & \\
BE20 & MAE, s & \ovalbox{17.0} & - & 62.3 & 65.5 \\\hline

\end{tabular}
\end{table}

\begin{table}[h!]\fontsize{9}{9.3}\selectfont
\caption{\gls{cas} above transition point for each aircraft type.}
\label{tab:stats_ias_above_av}
\centering \begin{tabular}{|c|c|c|c|c|c|}
\hline
AC Type & Statistic & Gaussian & GMM & NF & BADA \\
\hline
B738 & KS distance & 0.159 & \odashbox{0.143} & \odashbox{0.139} & \\
B738 & Wasserstein & 2.73 & 2.34 & \ovalbox{2.17} & \\
B738 & MAE & 0.429 & 0.326 & \ovalbox{0.169} & 14.7 \\\hline
A320 & KS distance & \odashbox{0.125} & 0.134 & \odashbox{0.127} & \\
A320 & Wasserstein & \ovalbox{1.74} & 1.91 & 1.86 & \\
A320 & MAE & \odashbox{0.170} & 0.186 & \odashbox{0.168} & 12.8 \\\hline
A319 & KS distance & \ovalbox{0.148} & 0.156 & 0.165 & \\
A319 & Wasserstein & \ovalbox{2.06} & 2.40 & 2.56 & \\
A319 & MAE & \ovalbox{0.589} & 0.640 & 0.835 & 14.5 \\\hline
DH8D & KS distance & 0.0886 & \ovalbox{0.0529} & 0.0696 & \\
DH8D & Wasserstein & 2.05 & \ovalbox{1.23} & 1.43 & \\
DH8D & MAE & 0.653 & \ovalbox{0.621} & 0.747 & 18.1 \\\hline
E190 & KS distance & \ovalbox{0.156} & 0.199 & 0.193 & \\
E190 & Wasserstein & \ovalbox{1.61} & 1.89 & 1.85 & \\
E190 & MAE & 0.228 & 0.272 & \ovalbox{0.188} & 2.80 \\\hline
B772 & KS distance & 0.137 & - & \ovalbox{0.126} & \\
B772 & Wasserstein & 2.22 & - & \ovalbox{1.76} & \\
B772 & MAE & 0.250 & - & \ovalbox{0.140} & 11.2 \\\hline
A388 & KS distance & \ovalbox{0.130} & - & 0.147 & \\
A388 & Wasserstein & 1.67 & - & \ovalbox{1.57} & \\
A388 & MAE & \ovalbox{0.521} & - & 0.749 & 5.80 \\\hline
E170 & KS distance & \odashbox{0.164} & - & \odashbox{0.160} & \\
E170 & Wasserstein & 2.12 & - & \ovalbox{1.94} & \\
E170 & MAE & \ovalbox{0.514} & - & 0.689 & 5.60 \\\hline
C56X & KS distance & 0.156 & 0.143 & \ovalbox{0.0981} & \\
C56X & Wasserstein & 4.14 & 3.53 & \ovalbox{1.44} & \\
C56X & MAE & 0.586 & \ovalbox{0.482} & 0.784 & 7.58 \\\hline
PC12 & KS distance & 0.162 & - & \ovalbox{0.118} & \\
PC12 & Wasserstein & 2.07 & - & \ovalbox{1.20} & \\
PC12 & MAE & 1.05 & - & \ovalbox{0.586} & 15.5 \\\hline
F2TH & KS distance & \ovalbox{0.302} & - & 0.360 & \\
F2TH & Wasserstein & \ovalbox{2.42} & - & 2.58 & \\
F2TH & MAE & 1.04 & - & \ovalbox{0.954} & 1.86 \\\hline
BE20 & KS distance & \odashbox{0.204} & - & \odashbox{0.200} & \\
BE20 & Wasserstein & 3.67 & - & \ovalbox{3.26} & \\
BE20 & MAE & \odashbox{1.77} & - & \odashbox{1.75} & 13.8 \\\hline

\end{tabular}
\end{table}

\begin{table}[h!]\fontsize{9}{9.3}\selectfont
\caption{\gls{cas} below transition point for each aircraft type.}
\label{tab:stats_ias_below_av}
\centering \begin{tabular}{|c|c|c|c|c|c|}
\hline
AC Type & Statistic & Gaussian & GMM & NF & BADA \\
 \hline
B738 & KS distance & 0.149 & 0.0693 & \ovalbox{0.0604} & \\
B738 & Wasserstein & 4.49 & 2.51 & \ovalbox{2.01} & \\
B738 & MAE & 1.11 & 0.703 & \ovalbox{0.577} & 21.7 \\\hline
A320 & KS distance & 0.0932 & \ovalbox{0.0580} & 0.0614 & \\
A320 & Wasserstein & 2.57 & \odashbox{1.57} & \odashbox{1.61} & \\
A320 & MAE & \ovalbox{0.176} & 0.370 & 0.204 & 16.5 \\\hline
A319 & KS distance & 0.110 & \ovalbox{0.0638} & 0.0716 & \\
A319 & Wasserstein & 3.05 & \ovalbox{1.81} & 2.17 & \\
A319 & MAE & \ovalbox{0.413} & 0.696 & 1.26 & 17.4 \\\hline
DH8D & KS distance & 0.103 & \odashbox{0.0815} & \odashbox{0.0825} & \\
DH8D & Wasserstein & 2.73 & \ovalbox{1.64} & 2.05 & \\
DH8D & MAE & 1.34 & 1.02 & \ovalbox{0.0496} & 30.5 \\\hline
E190 & KS distance & 0.184 & \odashbox{0.103} & \odashbox{0.104} & \\
E190 & Wasserstein & 4.2 & \odashbox{1.62} & \odashbox{1.58} & \\
E190 & MAE & 0.549 & \ovalbox{0.254} & 0.516 & 2.61 \\\hline
B772 & KS distance & \ovalbox{0.0851} & - & 0.0896 & \\
B772 & Wasserstein & 2.80 & - & \ovalbox{2.23} & \\
B772 & MAE & 0.677 & - & \ovalbox{0.243} & 12.7 \\\hline
A388 & KS distance & \odashbox{0.0893} & \odashbox{0.0877} & - & \\
A388 & Wasserstein & 3.16 & \ovalbox{2.48} & - & \\
A388 & MAE & 0.777 & \ovalbox{0.559} & - & 2.97 \\\hline
E170 & KS distance & 0.239 & - & \ovalbox{0.156} & \\
E170 & Wasserstein & 4.97 & - & \ovalbox{2.89} & \\
E170 & MAE & 1.09 & - & \ovalbox{0.661} & 20.9 \\\hline
C56X & KS distance & 0.135 & 0.0899 & \ovalbox{0.0722} & \\
C56X & Wasserstein & 3.37 & 2.37 & \ovalbox{1.93} & \\
C56X & MAE & \ovalbox{0.414} & 0.496 & 0.951 & 10 \\\hline
PC12 & KS distance & 0.209 & - & \ovalbox{0.0648} & \\
PC12 & Wasserstein & 3.69 & - & \ovalbox{1.18} & \\
PC12 & MAE & 2.13 & - & \ovalbox{1.11} & 20.5 \\\hline
AT76 & KS distance & \ovalbox{0.170} & 0.189 & 0.191 & \\
AT76 & Wasserstein & \ovalbox{2.91} & 3.16 & 3.22 & \\
AT76 & MAE & \ovalbox{1.98} & 2.12 & 2.16 & 5.31 \\\hline
F2TH & KS distance & \odashbox{0.230} & - & \odashbox{0.223} & \\
F2TH & Wasserstein & \ovalbox{5.03} & - & 5.25 & \\
F2TH & MAE & 0.942 & - & \ovalbox{0.805} & 7.07 \\\hline
BE20 & KS distance & 0.126 & - & \ovalbox{0.0762} & \\
BE20 & Wasserstein & 3.39 & - & \ovalbox{1.88} & \\
BE20 & MAE & \ovalbox{0.827} & - & 0.915 & 23.0 \\\hline

\end{tabular}
\end{table}

\begin{table}[h!]\fontsize{9}{9.3}\selectfont
\caption{\gls{rocd} above transition point for each aircraft type.}
\label{tab:stats_rocd_above_av}
\centering \begin{tabular}{|c|c|c|c|c|c|}
\hline
AC Type & Statistic & Gaussian & GMM & NF & BADA \\
\hline
B738 & KS distance & 0.145 &\odashbox{0.0972} & \odashbox{0.0945} & \\
B738 & Wasserstein & 180 & \odashbox{141} & \odashbox{137} & \\
B738 & MAE & \odashbox{24.5} & 47.2 & 29.8 & 838 \\\hline
A320 & KS distance & 0.120 & \odashbox{0.113} & \odashbox{0.116} & \\
A320 & Wasserstein & \odashbox{180} & \odashbox{185} & \odashbox{182} & \\
A320 & MAE & \ovalbox{66.0} & 77.4 & 74.4 & 972 \\\hline
A319 & KS distance & 0.0887 & \ovalbox{0.0827} & 0.0925 & \\
A319 & Wasserstein & \odashbox{140} & \odashbox{137} & 150 & \\
A319 & MAE &\odashbox{66.6} & \odashbox{66.5} & \odashbox{64.4} & 852 \\\hline
DH8D & KS distance & 0.109 & \ovalbox{0.103} & 0.111 & \\
DH8D & Wasserstein & 108 & \ovalbox{87.9} & 97.0 & \\
DH8D & MAE & \odashbox{42.5} & 45.6 & \odashbox{42.4} & 247 \\\hline
E190 & KS distance & 0.083 & \odashbox{0.0664} & \odashbox{0.0652} & \\
E190 & Wasserstein & 93 & \odashbox{72.1} & \odashbox{75.5} & \\
E190 & MAE & 43.1 & \ovalbox{36.0} & 41.7 & 1100 \\\hline
B772 & KS distance & 0.145 & - & \ovalbox{0.105} & \\
B772 & Wasserstein & 176 & - & \ovalbox{141} & \\
B772 & MAE & 76.0 & - & \ovalbox{57.7} & 937 \\\hline
A388 & KS distance & 0.165 & - & \ovalbox{0.149} & \\
A388 & Wasserstein & 190 & - & \ovalbox{160} & \\
A388 & MAE & 82.3 & - & \ovalbox{46.7} & 478 \\\hline
E170 & KS distance & \ovalbox{0.110} & - & 0.121 & \\
E170 & Wasserstein & \ovalbox{109} & - & 120 & \\
E170 & MAE & \odashbox{51.7} & - & \odashbox{51.3} & 2120 \\\hline
C56X & KS distance & \odashbox{0.110} & \odashbox{0.107} & 0.156 & \\
C56X & Wasserstein & 80.9 & \ovalbox{71.5} & 90.1 & \\
C56X & MAE & \odashbox{23.7} & \odashbox{24.5} & 61.8 & 968 \\\hline
PC12 & KS distance & 0.211 & - & \ovalbox{0.145} & \\
PC12 & Wasserstein & 119 & - & \ovalbox{77.2} & \\
PC12 & MAE & 71.9 & - & \ovalbox{24.9} & 982 \\\hline
F2TH & KS distance & \ovalbox{0.165} & - & 0.194 & \\
F2TH & Wasserstein & \ovalbox{156} & - & 199 & \\
F2TH & MAE & \ovalbox{72.6} & - & 109 & 1640 \\\hline
BE20 & KS distance & \ovalbox{0.175} & - & 0.192 & \\
BE20 & Wasserstein & \odashbox{122} & - & \odashbox{124} & \\
BE20 & MAE & 84.2 & - & \ovalbox{50.5} & 231 \\\hline
\end{tabular}
\end{table}

\begin{table}[h!]\fontsize{9}{9.3}\selectfont
\caption{\gls{rocd} below transition point for each aircraft type.}
\label{tab:stats_rocd_below_av}
\centering \begin{tabular}{|c|c|c|c|c|c|}
\hline
AC Type & Statistic & Gaussian & GMM & NF & BADA \\
 \hline
B738 & KS distance & 0.149 & 0.0493 & \ovalbox{0.041} & \\
B738 & Wasserstein & 187 & \ovalbox{57.9} & 67.8 & \\
B738 & MAE & 33.7 & \ovalbox{12.2} & 24.1 & 142 \\\hline
A320 & KS distance & 0.112 & \ovalbox{0.0306} & 0.0340 & \\
A320 & Wasserstein & 126 & \ovalbox{41.7} & 49.1 & \\
A320 & MAE & \ovalbox{13.1} & 15.8 & 16.0 & 200 \\\hline
A319 & KS distance & 0.121 & \odashbox{0.0407} & \odashbox{0.0411} & \\
A319 & Wasserstein & 130 & \ovalbox{51.2} & 62.7 & \\
A319 & MAE & \ovalbox{16.7} & 19.8 & 30.6 & 312 \\\hline
DH8D & KS distance & \odashbox{0.0849} & \odashbox{0.0879} & 0.100 & \\
DH8D & Wasserstein & 75 & \ovalbox{54.6} & 64.7 & \\
DH8D & MAE & 24.7 & \odashbox{9.15} & \odashbox{9.16} & 23.0 \\\hline
E190 & KS distance & 0.0764 & 0.0743 & \ovalbox{0.0700} & \\
E190 & Wasserstein & 74.3 & 67.9 & \ovalbox{63.8} & \\
E190 & MAE & 25.7 & 23.4 & \ovalbox{17.7} & 548 \\\hline
B772 & KS distance & 0.111 & - & \ovalbox{0.065} & \\
B772 & Wasserstein & 111 & - & \ovalbox{76.6} & \\
B772 & MAE & 46.7 & - & \ovalbox{22.1} & 118 \\\hline
A388 & KS distance & 0.162 & - & \ovalbox{0.0678} & \\
A388 & Wasserstein & 183 & - & \ovalbox{87.5} & \\
A388 & MAE & 65.4 & - & \ovalbox{41.0} & 239 \\\hline
E170 & KS distance & 0.115 & - & \ovalbox{0.103} & \\
E170 & Wasserstein & 104 & - & \ovalbox{93.7} & \\
E170 & MAE & 62.1 & - & \ovalbox{35.3} & 1380 \\\hline
C56X & KS distance & 0.099 & \ovalbox{0.0925} & 0.117 & \\
C56X & Wasserstein & 82.4 & \ovalbox{67.7} & 77.7 & \\
C56X & MAE & \odashbox{33.3} & \odashbox{34.2} & 42.2 & 513 \\\hline
PC12 & KS distance & 0.130 & - & \ovalbox{0.0972} & \\
PC12 & Wasserstein & 89.2 & - & \ovalbox{46.2} & \\
PC12 & MAE & 62.5 & - & \ovalbox{6.50} & 788 \\\hline
AT76 & KS distance & 0.140 & 0.127 & \ovalbox{0.120} & \\
AT76 & Wasserstein & 71.8 & 65.8 & \ovalbox{62.1} & \\
AT76 & MAE & 51.5 & 46.1 & \ovalbox{38.3} & 667 \\\hline
F2TH & KS distance & \ovalbox{0.104} & - & 0.110 & \\
F2TH & Wasserstein & \ovalbox{90.6} & - & 106 & \\
F2TH & MAE & 52.5 & - & \ovalbox{31.1} & 1020 \\\hline
BE20 & KS distance & 0.209 & - & \ovalbox{0.126} & \\
BE20 & Wasserstein & 177 & - & \ovalbox{82.7} & \\
BE20 & MAE & 151 & - & \ovalbox{25.9} & 47.8 \\\hline

\end{tabular}
\end{table}

\end{document}